\documentclass[apj]{emulateapj}
\usepackage[colorlinks,linkcolor={blue},citecolor={blue},urlcolor={red}]{hyperref}
\usepackage{mathtools}
\usepackage{epsfig,graphicx,natbib,amsmath,amsfonts,amssymb}%,xfrac}
\usepackage{color}
\usepackage{multirow}
\usepackage{booktabs}
\usepackage{amssymb}
\usepackage{threeparttablex} 
\usepackage{booktabs} 
\usepackage{mathrsfs}  

\usepackage{longtable}
\usepackage{array} % for extrarowheight
\usepackage[dvipsnames]{xcolor}  
\usepackage{lineno}

\newcommand{\msolar}{${\rm M}_\odot$}

\newcommand{\myemail}{\email{enci.wang@phys.ethz.ch}}

\shorttitle{Similar signatures of co-planar inflow and disk warps}
\shortauthors{Wang \& Lilly}

\graphicspath{{fig/}}
\defcitealias{Wang-22}{Paper I}

\begin{document}

%\title {Coplanar gas inflow can be hidden within warped galactic gas disks of  low-to-intermediate inclinations}

\title {The similar signatures of coplanar gas inflow and disk warps in galactic gas kinematic maps}

\author {Enci Wang\altaffilmark{1},
Simon J. Lilly\altaffilmark{1}} \myemail

\altaffiltext{1}{Department of Physics, ETH Zurich, Wolfgang-Pauli-Strasse 27, CH-8093 Zurich, Switzerland}

\begin{abstract}  

Hydrodynamic simulations suggest that galactic gas disks form when coplanar gas spirals into the inner regions of the disk.  We recently presented a simple ``modified accretion disk" model of viscous galactic disks in which star-formation is fed by a radial flow of gas.  However, little observational evidence has been presented for such inflows, which are expected to be only a few km s$^{-1}$ in the central regions of the disk, i.e. within three disk scale-lengths, but could reach of order 50-100 km s$^{-1}$ in the very outer disk.   The effects of systematic inflow on the 2-d velocity field are examined and it is shown that these are quite similar to those produced by geometric warps of the disks, with twist distortions of {\it both} the kinematic major and minor axes. This makes it potentially difficult to distinguish between these in practice.  By comparing the handedness of the observed twisting of the kinematic axes and of the spiral arms for a sample of nearby galaxies, we find (assuming that the spiral arms are generally trailing) that the effects of warps are in fact likely to dominate over the effects of radial inflows.   However, the common practice of treating these twist distortions of the kinematic major and minor axes as being due {\it only} to warps can lead, for galaxies of low-to-intermediate inclinations, to substantial underestimates of any systematic inflow.

\end{abstract}


\section{Introduction}
\label{sec:introduction}

It is well known that cold gas accretion is required to sustain star formation and size growth during the evolution of star-forming (SF) galaxies.  The existing cold gas in SF galaxies can only sustain the star formation for a few billion years or less \citep{Binney-00, Bigiel-11, Wang-13, Madau-15}. This indicates the need for gas accretion to continually replenish the fuel for star formation \citep{Fraternali-12}. 

Galaxies can accrete cold gas via capture of gas-rich dwarf galaxies and also through smooth accretion \citep[e.g.][]{Lacey-94, Murali-02, Sancisi-08, Keres-09, Bouche-10, LHuillier-12, Conselice-13, Sanchez-Almeida-14, Rodriguez-Gomez-15}.  The former may be linked with the presence of gas tails \citep{Kregel-01, Oosterloo-10} and lopsided HI morphologies \citep{Shang-98, Thilker-07}. 
However, the estimation of neutral gas accretion rate by the high velocity neutral-gas clouds is a factor of 3-5 less than the observed star formation rate (SFR) for M31/Milky Way-type galaxies \citep{Putman-06, Richter-12}. This suggests the importance of the smooth gas accretion, but convincing  evidence of this smooth accretion from observations is still lacking, and the geometry of such accretion is still correspondingly uncertain. 

Hydrodynamical simulations can be a powerful tool to understand the accretion of gas. Multi-zoom cosmological simulations suggest that smooth accretion dominates the flow of gas, while mergers may be important for the most massive high-redshift galaxies \citep{Murali-02, Keres-09, LHuillier-12}. Gas accretion along filamentary streams that avoid being shock-heated in the outer halo is known as ``cold-mode'' accretion. This is thought to occur in low mass galaxies at high-redshift \citep{Keres-05, Dekel-06, Ocvirk-08, Brooks-09, vandeVoort-11, Stern-20}.  From more massive disk galaxies, the accretion from the cooling of hot halo gas may be more important \citep{Keres-05, Ocvirk-08, Nelson-13, Stern-20}. 

Multiple simulations based on different hydrodynamical codes and carried out by different research groups show that the inflowing gas is almost co-planar and more or less co-rotating with the gas disk, at least at low redshifts ($z<0.8$), regardless of its thermal history \citep[e.g.][]{Keres-05, Stewart-11, Danovich-15, Stewart-17, Peroux-20, Trapp-21, Hafen-22, Gurvich-22}. In contrast, the outflow of gas that is driven by stellar winds and/or supernova (SN) explosions is preferentially along the direction that is perpendicular to the disk \citep[e.g.][]{Nelson-19, Peroux-20, Trapp-21}.  These results are also consistent with the structure and kinematic of circumgalactic medium (CGM), which are traced by the Mg II absorption from the observations \citep{Bordoloi-11, Bouche-12, Kacprzak-12, Diamond-Stanic-16, Bielby-17, Tumlinson-17, Peroux-17, Schroetter-19}. 

Motivated by these simulations, we constructed in previous papers a disk formation model that simply treats the gas disk of SF galaxies as an idealized ``modified'' (or ``leaking") accretion disk \citep[MAD;][hereafter \citetalias{Wang-22}]{Wang-22}.  In contrast to the classical accretion disks found around compact objects, the gas inflow in the MAD galactic disk gradually decreases towards the galactic center as gas is consumed by star formation or removed from the disk by any associated outflows.  

In \citetalias{Wang-22}, we investigated the possible mechanisms of how coplanar inflow could happen within a gas disk in terms of viscous processes within the disk. We showed that magneto-rotational instability is an attractive and plausible source of viscosity for the transport of mass inwards and angular momentum outwards within the gas disk. We showed that magnetic viscosity could naturally, under certain circumstances, produce the observed exponential profiles of SFR surface density \citep[e.g.][]{Wang-19} in galactic disks. However, two obvious questions raised by the simple MAD model (and actually independent of the source of the viscosity in the disk) concerned whether this idea was consistent with the observed metallicity profiles and with the observed kinematic maps of galactic gas disks.

In the second paper of this series \citep{Wang-22b}, we therefore examined the question of metallicity and found that the MAD model could account for the radial profiles of gas-phase metallicity for nearby galaxies. Not least, a negative gradient of gas-phase metallicity with radius is a natural consequence of the progressive enrichment of the gas by in-situ star formation as it flows inwards through the disk. 

In the present work, the third paper of this series, we examine the kinematic properties of the inflowing gas expected in the MAD framework. We find that the velocity of the required coplanar inflow is only a few km s$^{-1}$ at small radii ($<3h_{\rm R}$),  but gradually increases with radius and can reach 50-100 km s$^{-1}$ in the very outskirts of the gas disk. Interestingly, based on IllustrisTNG-100 simulations \citep{Nelson-18}, \cite{SWang-22} have recently found that the radial velocity of inflowing gas increases with radius for SF galaxies in the simulation, and reaches $\sim$55 km s$^{-1}$ at a radius of 10 times of half-mass radius \citep[see also][]{Trapp-21, Hafen-22}.  The good consistency between the MAD model and hydrodynamical simulations is not surprising given that the idealized MAD model was motivated by the results of such simulations. 

In contrast, however, by adopting a Fourier decomposition scheme to ten THINGS \citep[The HI Nearby Galaxy Survey;][]{Walter-08} galaxies, \cite{Schmidt-16} found no, or at best only very weak, radial velocities ($\sim$ 10 km s$^{-1}$ or less) in the neutral hydrogen, even at the outskirts of the disk. More recently, \cite{DiTeodoro-21} extracted the radial motion for 54 local disk galaxies from 21 cm emission datacubes. They also claimed that most galaxies show only very small radial velocities of a few km s$^{-1}$, positive and negative, throughout their HI disks, with no evidence of any {\it systematic} radial inflow and without any clear indications of an increase in flow velocity in the outermost regions \citep[e.g.][]{Warner-73, Trachternach-08, Speights-19, Bisaria-22}.

This apparent inconsistency between these observational results and the expectations of both our MAD model and, more generally, the large body of hydrodynamical simulations, could be due to a few different reasons. First, the inflowing gas with significant radial velocities in the outermost parts of the gas disks may not be neutral.  In the FIRE simulation, \cite{Hafen-22} found hot gas at the virial-temperature dominates the inflowing gas on scales greater than 20 kpc. On the other hand, \cite{SWang-22} found in the IllustrisTNG-100 simulation that the inflowing gas is cold out to 10-20 times the half-stellar-mass of galaxies.  Second, and related to the first, since the substantial radial inflows are predicted only at very large radii, the coverage of the current HI surveys may be not large enough to see significant radial inflow.  In this work, we explore another possibility, namely that the signatures of radial inflow could have been missed, at least partly, in analyses of the observed 2-d velocity fields because of the presence of geometric warps in the disks. 

%7. The organization of this work. 
The paper is organized as follows. In Section \ref{sec:2}, we will briefly review the MAD model, and give the prediction for the radial profile of the required radial inflow velocity.  We then in Section \ref{sec:3} explore the distortions that should be imprinted by such inflows on the 2-d projected line-of-sight (LOS) velocity fields of gas disks and compare these distortions with those expected from warping of the disks. The two distortion patterns are broadly similar. 
In Section \ref{sec:3.2}, we apply a novel geometric analysis that is designed to exploit symmetry arguments to distinguish between these two possibilities. This is based on inferring the sense of rotation of the disk from the twist of the spiral arms, which are assumed to be trailing.  This analysis actually suggests that, overall, the effect of warps on disk kinematics do indeed likely dominate over that of inflows.  Motivated by this, in Section \ref{sec:4}, we then construct mock datacubes and examine whether composite disks, in which there are both (dominant) warps plus radial inflows of the required strength, can be distinguished from systems in which (as commonly done) it is assumed that warps alone are present.  This allows us to address the question of whether (or to what degree) significant radial inflows can be ``hidden'' within the distortions of the kinematic major and minor axes arising from the (uncertain) warped geometries of individual disks.  We then summarize this work in Section \ref{sec:5}. 

When quoting distance-dependent quantities, we adopt a flat cold dark matter cosmology model with $\Omega_m=0.27$, $\Omega_\Lambda=0.73$ and $H_0 = 70$ km s$^{-1}$Mpc$^{-1}$.

\section{The required co-planar gas inflow in the modified accretion disk model} \label{sec:2}

\subsection{The brief recap of the MAD model}

We have introduced and discussed the MAD model in \citetalias{Wang-22} and \cite{Wang-22b}. Here we only present a brief summary of the key assumptions of this disk evolution model. 

As mentioned in the Introduction, the MAD model is motivated by the gas exchange of galaxies in hydrodynamical simulations: co-planar inflow of gas dominates the gas accretion, while wind-driven outflows potentially leave perpendicular to the disk, especially for low redshift SF galaxies \citep[e.g.][]{Trapp-21, Hafen-22}. The dominance of the radial inflow is the most important assumption in the MAD model.  For simplicity, in addition to the co-planar inflow, we combine any ex-planar inflows with the expected ex-planar outflow (driven by stellar feedback) into an ``effective outflow" and assume that this is always proportional to the  instantaneous SFR surface density ($\Sigma_{\rm SFR}$) multiplied by an effective mass-loading factor $\lambda$, which can be positive or mildly negative ($\lambda \gg -1$).  Also for simplicity, this $\lambda$ is assumed to be the same at all radii of the disk. 

We also assume that the local $\Sigma_{\rm SFR}$ in the disk is instantaneously determined by the local gas surface density ($\Sigma_{\rm gas}$) via some ``star formation law" expressed in terms of a star formation efficiency (SFE=$\Sigma_{\rm gas}/\Sigma_{\rm SFR}$).  The local gas surface density is regulated by the interplay between the coplanar gas inflow, and the removal of gas by star formation and any associated outflows \citep[e.g.][]{ Sommer-Larsen-89, Thon-98, Bouche-10, Schaye-10, Lilly-13}. In other words, the gas disk can be treated as a ``gas regulator system"  \citep[e.g.][]{Lilly-13, Wang-19, Wang-21} at all galactocentric radii. 

After star formation occurs, a fraction of the mass of the newly formed stars would be subsequently returned to the interstellar medium through stellar winds and supernova explosions. We denote this return fraction as $R$ and adopt the instantaneous recycling approximation. We take $R =$ 0.4 from the \cite{Chabrier-03} initial mass function (IMF) and stellar population models \citep[][]{Bruzual-03, Vincenzo-16}. 

Based on the above assumptions, the simple continuity equations for the conservation of the cold gas mass, the mass of metals, and the angular momentum are easily constructed (see equation 1 and 3 in \citetalias{Wang-22} and equation 2 in \cite{Wang-22b}).   With the additional assumption of a steady-state of the gas disk and adopting from observations some $\Sigma_{\rm SFR}$ profile - e.g. an exponential profile \citep[][\citetalias{Wang-22}]{Bigiel-08, Wyder-09, Gonzalez-Lopezlira-12, Gonzalez-Delgado-16, Casasola-17, Wang-19} - it is then straightforward to obtain the analytical solution of the inflow rate $\Phi(r)$, the gas-phase metallicity $Z(r)$, and the viscous stress required to maintain the $\Sigma_{\rm SFR}$. 

In \citetalias{Wang-22} we found that the most plausible source of the viscosity in the accretion disk that is required in order to drive the gas inflow through the disk is magneto-rotational instability \citep[MRI;][]{Balbus-91}. Therefore, we explored a disk evolution model that is driven by magnetic stresses.  This successfully produces the exponential profile of $\Sigma_{\rm SFR}$ with reasonable scalelength, regardless of the initial conditions of the gas disk (\citetalias{Wang-22}), provided that there is a connection between the local magnetic field and the $\Sigma_{\rm SFR}$ of the form that has been seen observationally ($B_{\rm tot}\propto \Sigma_{\rm SFR}^{0.15}$). 
The MRI-MAD combination may provide a  solution for the long puzzling origin of the exponential form of star-forming disks in galaxies. 

One of the striking features of this model is that the exponential form of $\Sigma_{\rm SFR}(r)$ does not depend on the assumed form of the star-formation law linking $\Sigma_{\rm SFR}$ and $\Sigma_{\rm gas}$, since the system will adjust to ensure the steady-state $\Sigma_{\rm SFR}(r)$.   This is in fact a generic feature of gas-regulator models.

In addition, the resulting metallicity profile in the MAD model has a simple analytic form \citep[see equation 12 in][]{Wang-22b}, which is solely determined by the $\Sigma_{\rm SFR}$ profile (e.g. the exponential scalelength of $h_{\rm R}$), the mass-loading factor, and the metallicity of the gas at the outer boundary.  It is, again, quite independent of the assumed SFE.  The solution of $Z(r)$ matches quite well the broad features of the extended metallicity profiles that are observed in nearby SF galaxies \citep{Wang-22b}.  These all strengthen the idea that the simple MAD model is likely to closely approximate reality. 

In this work, we focus on the radial motion of gas that is at the heart of the MAD model.  We note that the detailed assumptions of chemical enrichment and rotational support of the gas disk that were adopted in \citetalias{Wang-22} and \cite{Wang-22b} are not required for deriving the radial velocity of coplanar inflow in this paper, and can be relaxed.

\subsection{The inflow velocity required by the MAD model}

\begin{figure}
  \begin{center}
    \epsfig{figure=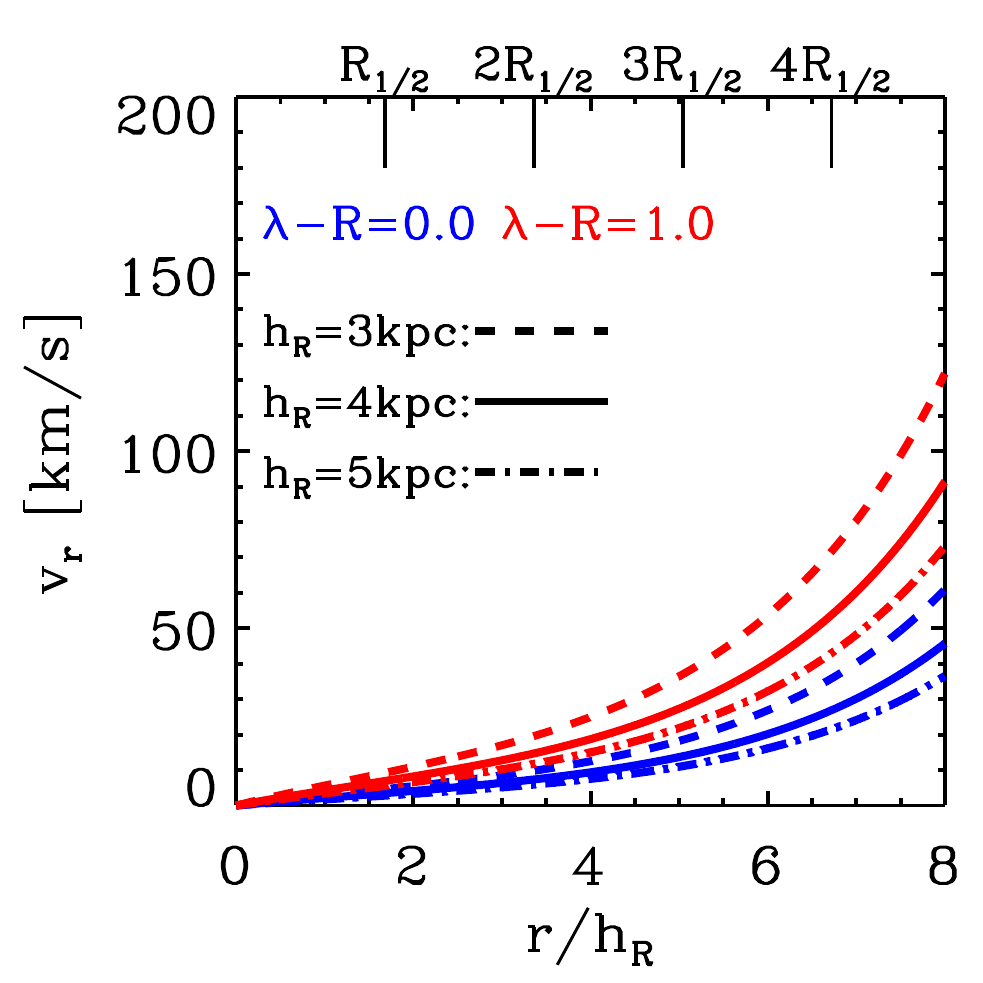,clip=true,width=0.45\textwidth}
    \end{center}
 \caption{ The required radial inflow velocity as a function of radius for a typical disk galaxy of $M_*=3\times10^{10}$\msolar\ in the MAD framework. Different colors indicate two different mass-loading factors, $\lambda$, and the three different line-styles indicate different exponential scale-lengths assumed for this galaxy. The $h_{\rm R}$-dependence of the radial inflow velocity follows from Equation \ref{eq:1}, provided that the assumed $\Sigma_{\rm gas}$ is only a function of $r/h_{\rm R}$ (see Equation \ref{eq:4}). }
  \label{fig:1}
\end{figure}

\begin{figure*}
  \begin{center}
    \epsfig{figure=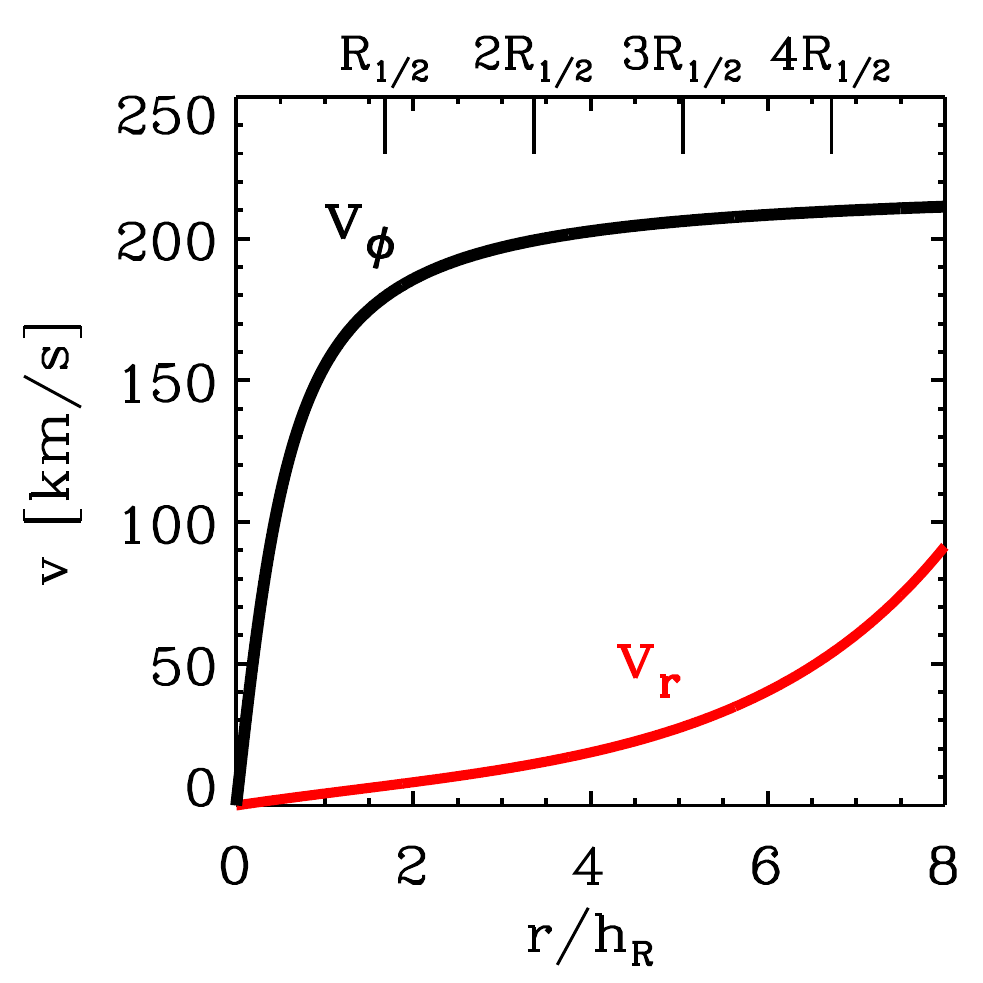,clip=true,width=0.45\textwidth}
    \epsfig{figure=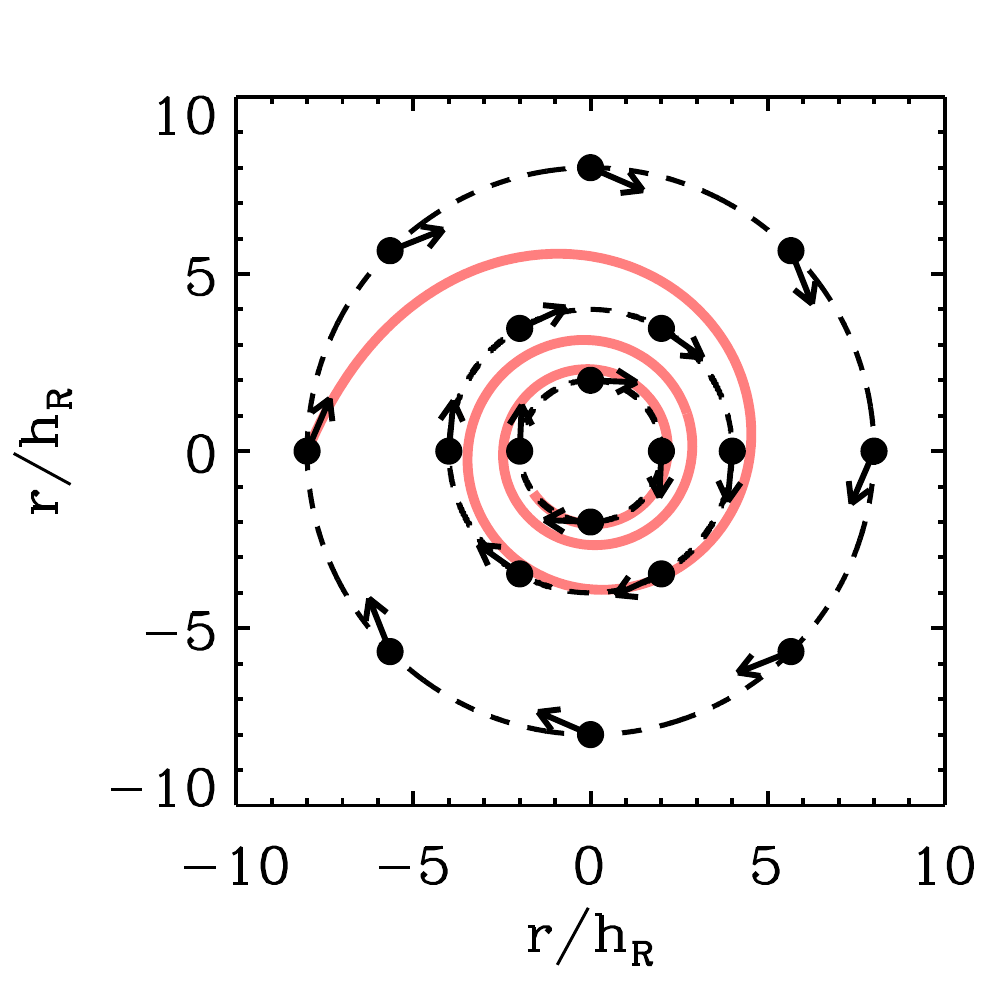,clip=true,width=0.45\textwidth}
    \end{center}
  \caption{
  Left panel: The circular velocity (black curve) and radial velocity (red curve) of the typical Main Sequence galaxy in Figure \ref{fig:1} but for only the case of $h_{\rm R}$=4 kpc and $\lambda-R$=1.  Right panel: the face-on velocity field (black arrows) and the inspiral path between 8$h_{\rm R}$ and 2$h_{\rm R}$ for a given gas element in this same galaxy (red helix). The dashed circles indicate the radii of 2$h_{\rm R}$, 4$h_{\rm R}$ and 8$h_{\rm R}$.  }
  \label{fig:2}
\end{figure*}

The radial velocity of co-planar inflowing gas can be written as: 
\begin{equation} \label{eq:1}
    v_{\rm r} = \frac{\Phi(r)}{2\pi r \Sigma_{\rm gas}(r)}. 
\end{equation}
In order to obtain the amplitude of $v_r$, the $\Sigma_{\rm gas}(r)$ and the radial inflow rate $\Phi(r)$ are needed.  

In \citetalias{Wang-22} it was shown that, with the assumption of an exponential $\Sigma_{\rm SFR}$, the steady-state solution for the radial inflow rate can be written as: 
\begin{equation} \label{eq:2}
     \frac{\Phi(r)}{1-R+\lambda} = {\rm SFR} \cdot [1+\eta-(x+1)\cdot \exp(-x)], 
\end{equation}
where $x$ is a scaled radius defined as $x=r/h_{\rm R}$, the SFR refers to the integrated SFR of the whole gas disk, and the factor $\eta$ accounts for any mass sink at the disk center, e.g. due to black hole accretion and associated jet-driven outflow.   

Unlike the solutions for $\Phi(r)$ and $Z(r)$, which depend {\it only} on the $\Sigma_{\rm SFR}(r)$ profile and the assumed $\lambda$ (and $\eta$), the radial velocity $v_{\rm r}(r)$ is sensitive to the $\Sigma_{\rm gas}$, i.e. effectively, given the input $\Sigma_{\rm SFR}(r)$, on the star formation law (or SFE) for the system.  In \citetalias{Wang-22}, we assumed for simplicity that the \cite{Kennicutt-98} star formation law applied everywhere, even though this star formation law is known to fail at the outskirts of gas disks \citep[e.g.][]{Leroy-08, Bigiel-10, Shi-11}.  In order to derive a more realistic $v_r(r)$ profile, we here adopt a more empirical $\Sigma_{\rm gas}(r)$ that is obtained directly from observations.  We use from \cite{Bigiel-12}: 
\begin{equation} \label{eq:3}
    \Sigma_{\rm gas} = 2.1 \Sigma_{\rm trans} \cdot e^{-\frac{r}{0.61R_{25}}}, 
\end{equation}
where $\Sigma_{\rm trans}$ is the gas surface density at the radius where the HI gas becomes dominant, and $R_{\rm 25}$ is defined as the 25 mag arcsec$^{-2}$ B-band isophote \citep{Schruba-11}. 
The $\Sigma_{\rm trans}$ does not vary greatly from galaxy to galaxy, and has a typical value of 14 ${\rm M_{\odot}}$ pc$^{-2}$ \citep{Leroy-08, Bigiel-08, Bigiel-12}. Adopting $R_{25}=3h_{\rm R}$ \citep{Casasola-17}, we can then rewrite Equation \ref{eq:3} as: 
\begin{equation} \label{eq:4}
    \Sigma_{\rm gas} = 29.4 \times e^{-\frac{r}{1.8h_{\rm R}}} \ [{\rm M_{\odot} pc^{-2}}]
\end{equation}
We note that the $h_{\rm R}$ is the scalelength of star-formation $\Sigma_{\rm SFR}$, rather than of the stellar mass (or gas) in the disk. 

With the assumption of $\Sigma_{\rm gas}$ in Equation \ref{eq:4}, Figure \ref{fig:1} shows the required $v_r(r)$ profiles as a function of radius in the MAD model. We show a typical SF galaxy with a stellar mass of $M_*=3\times10^{10}$\msolar\ (cf. figure 3 in \citetalias{Wang-22}).  The integrated SFR of this typical galaxy is set to be 3.5 \msolar${\rm yr}^{-1}$ adopting the star formation main sequence from \cite{Lilly-16} at a redshift of zero \citep[also see][]{Noeske-07, Speagle-14, Renzini-15}. For illustration, we consider three different values of the scalelength of the star-forming disk, $h_{\rm R} =$ 3, 4, or 5 kpc, and two different values of the effective wind load factor $\lambda-R = $0 or 1. 

As shown, the $v_{\rm r}$ is very small in the inner parts of the disk: a few km s$^{-1}$ within 4$h_{\rm R}$, corresponding to the edge of stellar disk.  This is consistent with previous work that the radial velocity required by chemodynamical models are only of the order of a few km s$^{-1}$ \citep{Bilitewski-12, Pezzulli-16, Wang-22b}.  However, the $v_r$ increases rapidly beyond 4$h_{\rm R}$, and can be as large as 50-100 km s$^{-1}$ at 8$h_{\rm R}$, depending on the assumed mass-loading factor. This is because the density of gas continues to drop but the required inflow stays more or less the same, being determined by the integrated SFR and outflow within that radius, so the inflow velocities must increase.
This behaviour is consistent with the results of hydrodynamical simulations that the radial velocity of gas increases with radius and can reach 30-70 km s$^{-1}$ at 10 times the half-mass radius \citep{ SWang-22, Hafen-22}. 

For disk galaxies, the circular velocity of the gas disk typically increases rapidly in the inner regions, and then becomes flat out to the very outer regions of the gas disk \citep[e.g.][]{Courteau-97, de-Blok-08}. As in \citetalias{Wang-22}, we assume for simplicity that the circular velocity can be written in a simple {\tt arctan} form \citep{Willick-99, Miller-11}: 
\begin{equation} \label{eq:5}
    v_{\phi}(r) = V_{\rm cir} \cdot \frac{2}{\pi} {\rm arctan}(r/R_{\rm t}), 
\end{equation}
where $V_{\rm cir}$ is the maximum circular velocity, and $R_{\rm t}$ characterizes the transition radius between the rising and flat parts of the rotation curve.  For the typical SF galaxy mentioned above (see Figure \ref{fig:1}), we adopt $V_{\rm cir}$ = 220 km s$^{-1}$ and $R_{\rm t}=0.5h_{\rm R}$ \citep[][\citetalias{Wang-22}]{Miller-11}. 

The left panel of Figure \ref{fig:2} shows both the circular velocity (black curve) and the radial velocity (red curve) as a function of radius for our typical SF galaxy, adopting for definiteness $h_{\rm R}$=4 kpc and $\lambda-R$=1.  One can thus obtain an idea of the trajectory of the inspiraling gas elements, assuming the azimuthal velocity is always close to circular, which is shown in the red helix of the right panel of Figure \ref{fig:2}. 
As shown, the inflowing gas gradually spirals in towards the disk center. This is similar to the trajectories seen in the hydrodynamical simulations \citep[see figure 2 in][]{Hafen-22}.  For the specific case shown in Figure \ref{fig:2}, it takes $\sim$1 Gyr for a given gas element to move from 8$h_{\rm R}$ down to 2$h_{\rm R}$, which is of the same order of the gas depletion timescale \citep{Shi-11, Wang-19}. The inflow velocity becomes very significant with respect to the circular velocity in the outskirts of gas disk ($\sim 8h_{\rm R}$ or larger) and the motion of gas in the disk strongly deviates from circular motion. Hydrodynamical simulations also show that the motions of gas is no longer rotationally supported far beyond the stellar disk \citep{Trapp-21, Hafen-22}, where the gas particles fall towards the center with the conservation of their angular momentum.  

It should be appreciated that the MAD model is intentionally constructed to provide a very simple and analytically-tractable conceptual framework for consideration of a viscous disk.  It makes no statement about the physical state of the gas (ionized or neutral) nor about whether there may be different disk components, possibly moving with different radial velocities, at different heights above the disk \citep[e.g.][]{Fraternali-08}.  The radial inflow velocities discussed in this section should be understood as average velocities of all the gas in the loosely-defined ``disk" of the galaxy.  Furthermore, the model is much better constrained in the inner regions of the disk, within say 3 disk scale-lengths, where the bulk of star-formation and gas consumption occurs, than in the outer regions that are primarily feeding this inner zone, without significant gas consumption in-situ.  Indeed, as stressed in \citetalias{Wang-22}, the choice of exactly where to place the outer boundary of the viscous disk, i.e. the outer limit of applicability of the viscous MAD model, is not very important.

\section{The signatures of radial inflow in the 2-d projected velocity field} \label{sec:3}

In Section \ref{sec:2}, we showed that the inflow velocity required in the MAD picture can be very significant far beyond the stellar disk ($\sim$50-100 km s$^{-1}$). Such large radial motions are likely to significantly modify the observed LOS velocity field of gas disks relative to that of pure circular motion. 
In this section, we investigate the basic signatures of radial motion on the observed projected velocity fields.

\begin{figure*}
  \begin{center}
    \epsfig{figure=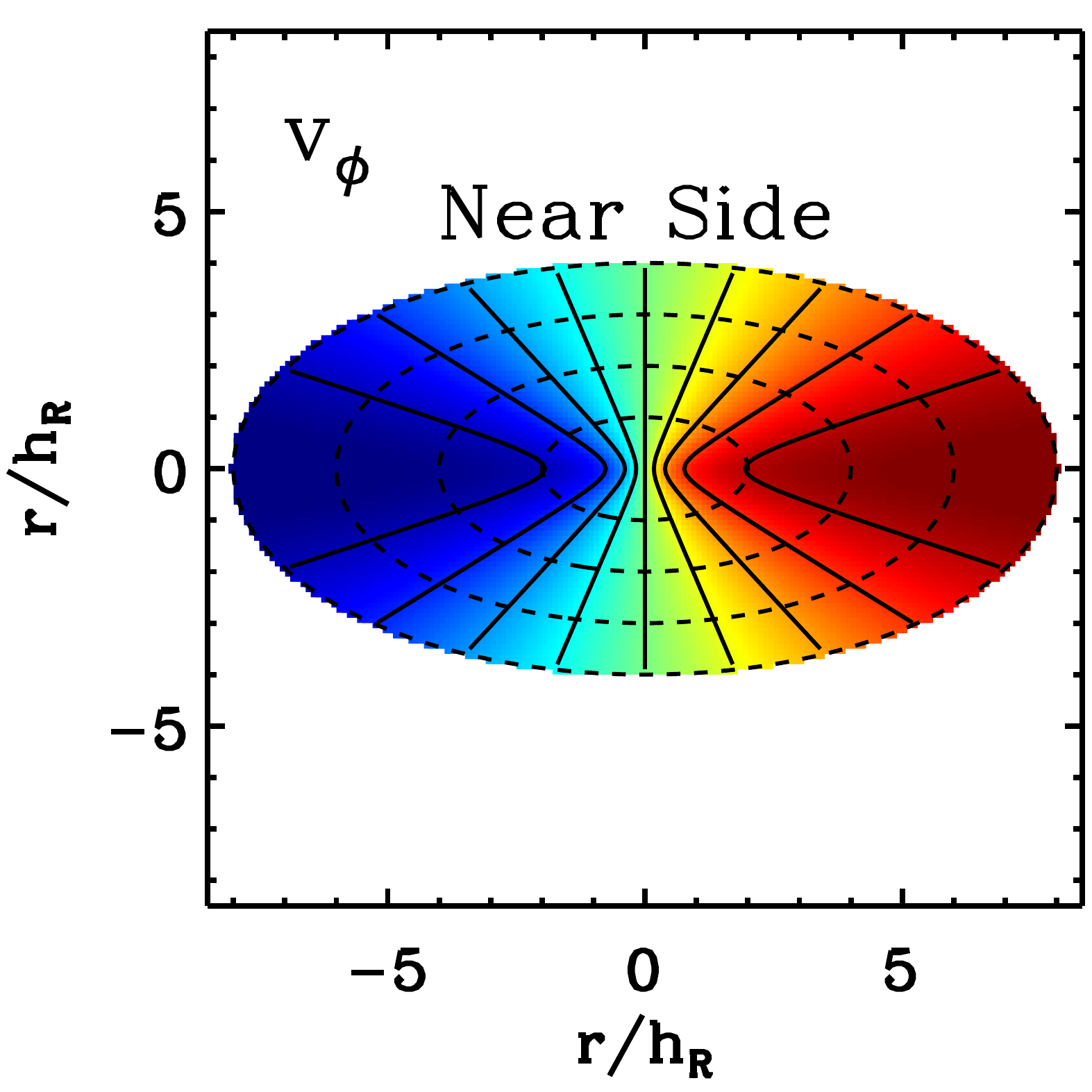,clip=true,width=0.305\textwidth}
    \epsfig{figure=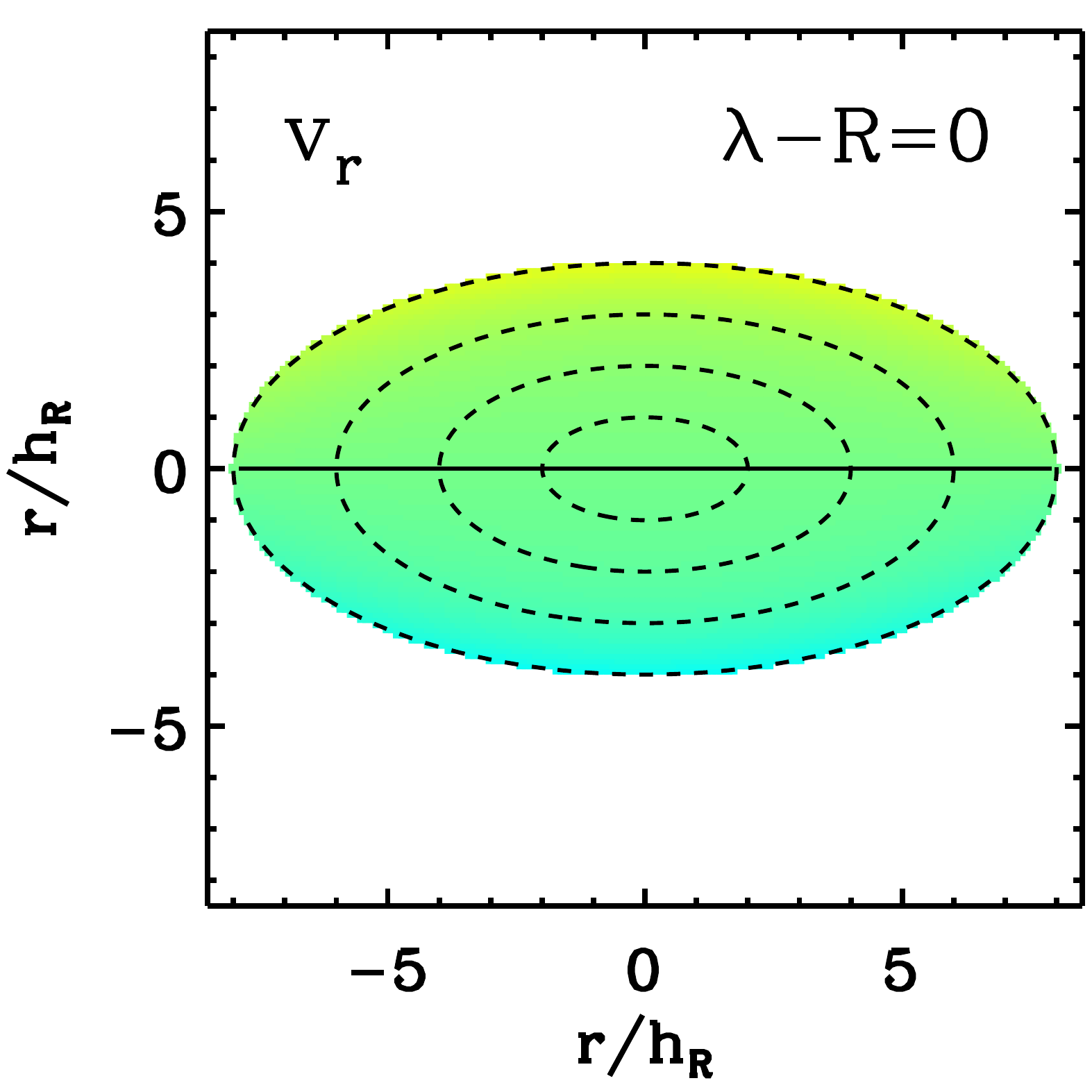,clip=true,width=0.305\textwidth}
    \epsfig{figure=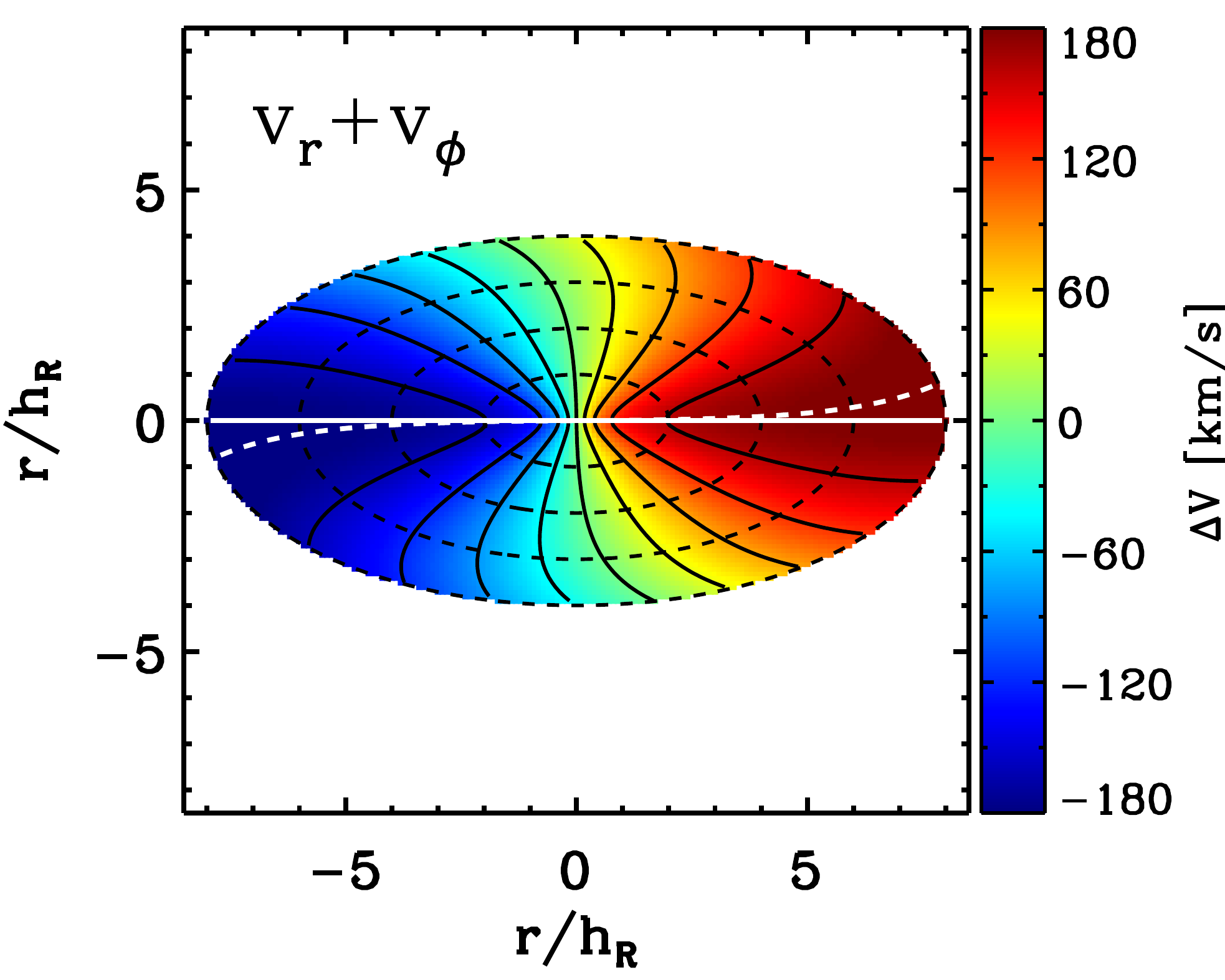,clip=true,width=0.38\textwidth}
    
    \epsfig{figure=./fig/kinematic/model/velocity_field_60_0-eps-converted-to.pdf,clip=true,width=0.305\textwidth}
    \epsfig{figure=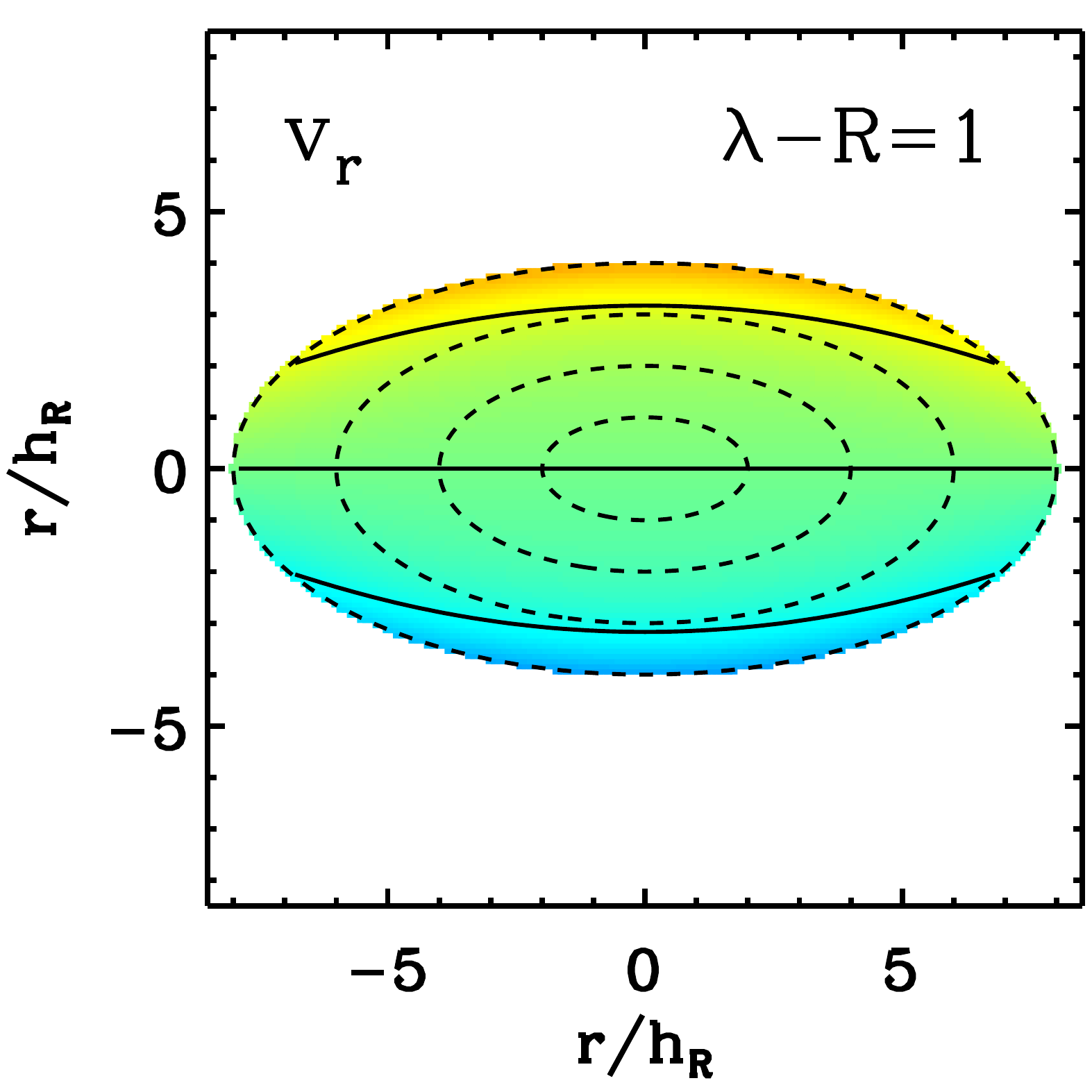,clip=true,width=0.305\textwidth}
    \epsfig{figure=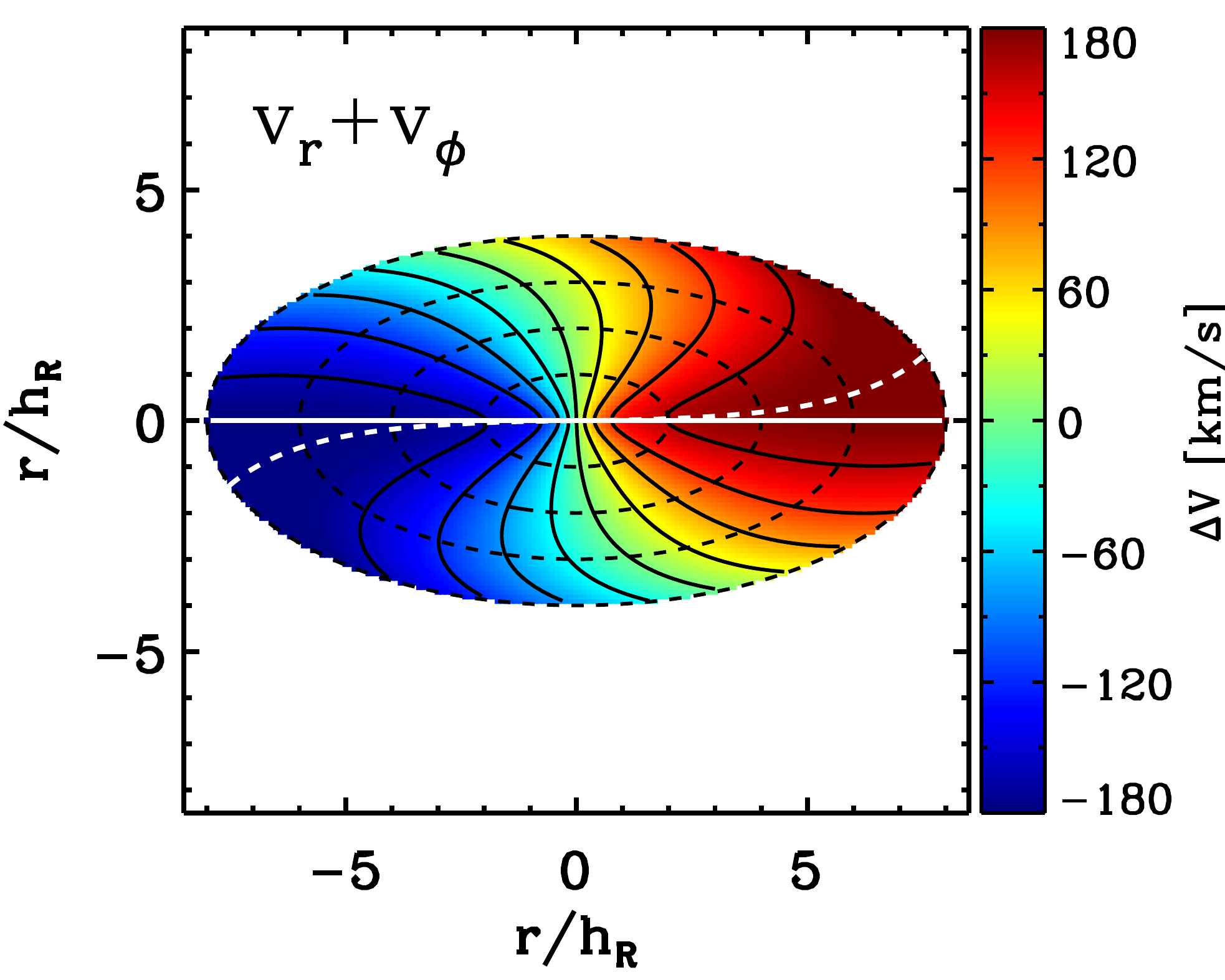,clip=true,width=0.38\textwidth}
    \end{center}
  \caption{The LOS velocity map of the typical Main Sequence galaxy shown in Figure \ref{fig:2} as seen at  an inclination (INC) of 60 degree.  The upper and lower row of panels assume $\lambda-R=0$ and $\lambda-R=1$, respectively (the latter requiring higher inflow velocities).  In each row, the leftmost panel shows the contribution of the circular motion, i.e. $v_{\phi}$, and the middle panel shows the contribution of the radial motion, i.e. $v_{\rm r}$. The rightmost panel shows the ``observed'' LOS velocity obtained by adding both $v_{\phi}$ and $v_{\rm r}$ components.
In the rightmost panels, the white solid line shows the geometric major axis of this galaxy, while the white dashed line shows the kinematic major axis, defined as the locus of minimum/maximum projected velocity at each radius.  
  In each panel, the black curves show iso-velocity contours with an interval of 40 km s$^{-1}$.  
  The dashed elliptical lines indicate radii of 2$h_{\rm R}$, 4$h_{\rm R}$, 6$h_{\rm R}$ and 8$h_{\rm R}$.
  }
  \label{fig:3}
\end{figure*}

Figure \ref{fig:3} shows the 2-d velocity field (projected onto the LOS) contributed by circular motion (left column), radial inflow (middle column), and the combination of the two (right column) for the same Main Sequence galaxy as above. The gas disk is assumed to be un-warped and to be seen with a (constant) inclination (INC) of 60 degrees, and to be rotating clockwise (in other words, the upper limb of the disk in each of the panels is closer to us than the lower limb). For illustration, we take $\lambda-R=0.0$ and $\lambda-R=1.0$ for the top and bottom row of panels.  These therefore differ by a factor of two in the required $v_r(r)$ profile.

As expected, the velocity field along the geometric major axis is only produced by the circular velocity component and that along the minor axis only by the radial velocity component. The velocity field with only circular motion is the classic ``spider diagram'' \citep[e.g.][and references therein]{Walter-08}, which is reflection-symmetric across the major axis.  The velocity field with only radial motion is reflection-symmetric across the minor axis.

When both components are added, the composite velocity field shows a centrosymmetric distortion relative to the velocity field obtained with purely circular motion.  It is noticeable that the kinematic major axis (white dashed line), defined to be the locus of the maximum projected velocity as a function of radius, is significantly deviated from the geometric major axis (white solid line) of the system, even though the radial velocities have, by definition, zero effect along the geometric major axis. Likewise, the kinematic minor axis, defined as the locus of zero projected velocity (relative to systemic), is similarly deviated from the geometric minor axis.  The kinematic major axis and the iso-velocity contours are systematically twisted in a coherent spiral pattern.  This deviation becomes more significant as the radial velocity increases towards the outer regions of the disk.   

It is easy to derive the amplitude of this deviation analytically. The LOS velocity with both circular $v_{\phi}(r)$ and radial $v_r(r)$ motions can be written as \citep[e.g.][]{Begeman-87}: 
\begin{equation} \label{eq:6}
    V_{\rm LOS} = V_{\rm sys} + {\tt sin}i \times (v_{\phi}{\tt cos}\theta + v_{\rm r} {\tt sin}\theta), 
\end{equation}
where $V_{\rm sys}$ is the systemic velocity of the galaxy, $i$ is the inclination angle (INC), and $\theta$ is the azimuthal angle {\it in the plane of the disk}, with $\theta=0^{\circ}$ corresponding to the geometric major axis. Here we only consider the very simple case of symmetric radial inflow and ignore other non-circular motions. We refer the reader to \cite{Schoenmakers-97} for more sophisticated models \citep[also see][]{ Wong-04, Spekkens-07, Sellwood-10}.

Based on Equation \ref{eq:6}, it is clear that the $V_{\rm LOS}$ reaches its extreme values relative to systemic velocity (i.e. the maximum and minimum observed values, defining the major kinematic axis) 
at
\begin{equation} \label{eq:7}
{\tt tan}\theta_{\rm major}= v_r/v_{\phi}.
\end{equation}
The corresponding extremal $V_{\rm LOS}$ can then be written as: 
\begin{equation} \label{eq:8}
    V_{\rm LOS}=  V_{\rm sys}  \pm  {\tt sin}i \cdot (v_{\phi}^2 + v_{r}^2)^{1/2}
\end{equation}
According to Equation \ref{eq:8}, disks with significant radial motions show larger LOS velocities (i.e. $(v_{\rm r}^2 + v_{\phi}^2)^{1/2}$) compared with their counterparts with purely circular motions.

Given the definition of $\theta$, the deviation of the kinematic major axis from the geometric major axis, $\Delta \theta_{\rm major}$, is identical to $\theta_{\rm major}$ in Equation \ref{eq:7}. Because $\theta$ was defined in the plane of the disk, we can then convert this $\Delta \theta_{\rm major}$ to the deviation of the kinematic and geometric major axes seen in the plane of the sky (i.e. as observed) denoting azimuthal angles in the plane of the sky with $\phi$:
\begin{equation} \label{eq:9}
 \Delta \phi_{\rm major} = {\tt arctan} (\frac{v_{\rm r} \cdot {\tt cos} i }{v_{\phi}} ). 
\end{equation}
The Equation \ref{eq:9} is therefore the analytic solution of the white dashed line (the kinematic major axis) in the two rightmost panels of Figure \ref{fig:3}. 

In the same way, we can calculate the deviation of the kinematic minor axis from the geometric minor axis.  The kinematic minor axis (denoted as $\theta_{\rm minor}$) is here defined to be the locus where the LOS velocity equals the systemic velocity, $V_{\rm sys}$. 
Since the geometric minor axis lies (by definition) at $\theta=90^{\circ}$, the deviation of the kinematic minor axis from the geometric minor axis, $\Delta \theta_{\rm minor}$, is just given by $\theta_{\rm minor}-90^{\circ}$. 
 
Letting $V_{\rm LOS}=V_{\rm sys}$, we can obtain the value of $\theta_{\rm minor}$ at the kinematic minor axis as: 
\begin{equation} \label{eq:10}
    {\tt tan} \theta_{\rm minor} = - v_{\phi}/v_r, 
\end{equation}
We can further obtain the deviation of the minor axis as: 
\begin{equation}
    {\tt tan}(\Delta \theta_{\rm minor}) =  v_{\rm r}/v_{\phi}. 
\end{equation}
In a similar way, we convert the $\Delta \theta_{\rm minor}$ of the disk plane to the one defined in the projected plane (or observed plane), which can be written as:
\begin{equation} \label{eq:12}
 \Delta \phi_{\rm minor} = {\tt arctan} (\frac{v_{\rm r}}{v_{\phi} \cdot {\tt cos} i}).
\end{equation}

It is worth stressing that the radial motion distorts both the minor and major kinematic axes of the LOS velocity fields from the geometric ones, but with different amplitudes.  The deviation of the minor axis (shown in Equation \ref{eq:12}) is larger than the deviation of the major axis (shown in Equation \ref{eq:9}), but this depends on the inclination of the disk. 

A geometric warp in the disk can also cause radial changes of the major and minor axes, due to the straightforward change of the geometry of the disk relative to the viewing direction \citep[e.g.][]{Jozsa-07,deBlok-08, Kamphuis-15, DiTeodoro-21}. However, the changes in the minor and major axes due to a warp of the disk should be identical. 

In principle this difference would enable these two signatures (due to radial velocities and due to a warp) to be distinguished.
However, for weakly inclined disks, the $\Delta \phi_{\rm major}$ produced by inflows is close to $\Delta \phi_{\rm minor}$. This implies that the effect of radial motions in weakly inclined disks will resemble the effect of a warp in the disk.  %, \textbf{which increases the difficulty to separate these two effects in practise}.  %producing a degeneracy between the two.

Due to this similarity of the effects of radial motion and warped disk, it is rather difficult to decompose the two based on the velocity fields of the HI gas. 
In principle, the isophotes of the column density map of gas could independently provide the geometric parameters of the disk, which potentially provides a way to decompose the above two effects. This approach may be applicable to some individual galaxies, such as NGC 5055 \citep{Battaglia-06}. However, the gas distribution is usually clumpy along with spiral arms, which strongly increases the difficulty of this approach. 

Recently, \cite{DiTeodoro-21} performed a three-step fitting method to try to isolate the radial motion using datacubes of HI emission.  In the first two steps, they fitted the warped geometry of the disk and the circular motion based on the location of the kinematic major axis. In the third step, they then fit the radial motion using the regions of the kinematic minor axis.  This method effectively attributes the radial distortion of the kinematic major axis to only the presence of a warp and ignores the distortion of the kinematic major axis that are instead caused by radial flows.   Based on the above discussion, this procedure will act to underestimate the radial motion on the disk \citep[see also][]{Schmidt-16} because it will explicitly ascribe the distortion of the kinematic major axis to the warp alone. As noted above, we stress again that radial flows do affect the location of the {\it kinematic} major axis, distorting it away from the geometric axis, even though radial flows make zero contribution to the velocity field along the {\it geometric} major axis itself.

It should also be noted that the measured circular velocity inferred from observations may also be affected by the possible presence of radial motions.  According to Equation \ref{eq:8}, the maximum/minimum observed LOS velocity is determined by both $v_{\phi}$ and $v_r$.  Ignoring possible radial motions in modelling 2-d velocity fields could potentially overestimate the amplitude of the circular motion, especially for the outer regions of the gas disk \citep[c.f.][]{Jones-99, deBlok-08, Kamphuis-15, Kam-17} where radial velocities may be high and, as shown below in Section \ref{sec:4.2.2}, for low inclination systems.

\section{A statistical test to determine whether warps or inflows are dominant} \label{sec:3.2}

In the previous section, we emphasized the difficulty of distinguishing the effects of radial flows from those associated with warps in disks in any individual system.  In this section, we develop a statistical test that allows us to determine whether, in the population as a whole, warps or radial inflows are the dominant effect. This is based on the breaking of symmetry that is implicit in requiring the radial flow to be inflow rather than outflow.

As shown in Figure \ref{fig:3}, the radial inflow has twisted the iso-velocity lines in a direction that is counter-clockwise as we move out in radius.  As noted above, the disks in Figure \ref{fig:3} are rotating clockwise (i.e. the upper limbs in the panels are closer to the observer than the lower limbs). Assuming that spiral arms in galaxies are generally trailing arms \citep[e.g.][]{Pasha-82, Iye-19}, then the spiral arms in this galaxy should also be twisted in a counter-clockwise direction moving out in radius.   It is easy to see that flipping the orientation of the disk, so that upper limb is now further away than the lower limb, would reverse the sense of the rotation as seen by the observer, and therefore reverse the twists of both the kinematic distortion and of the spiral arms.
If radial inflow is the dominant effect in distorting the iso-velocity contours of the galaxy, then we would therefore expect that the twists of the kinematic major axis and of the spirals arms should always be in the {\it same} sense, provided only that the spiral arms are always trailing. 

Warps are characterised by a migration of the rotation axis of the disk with radius. This could arise from a number of physical mechanisms. But in a symmetric universe, none of these should depend on the sense of rotation of the disk.  As a result, we would expect from symmetry arguments that the twists in the kinematic distortions of the velocity field due to warps should be uncorrelated with the sense of rotation and thus the twists of the spiral arms. We should as often see these two twists in opposite senses as in the same sense.

Examining the sense of the twists of the kinematic distortions and of the spiral arms (when visible) therefore provides a statistical test of whether radial inflow or warps dominate these distortions.  Provided that spiral arms are trailing, then, if warps dominate then we would expect a 50:50 split between galaxies with twists in the same sense and in the opposite sense, whereas if radial inflows dominate then these distortions should always be in the same sense.
Clearly the correlation would be weakened if some galaxies have leading spiral arms.

We therefore examined the HI velocity fields and the optical images for a sample of 54 nearby galaxies taken from \cite{DiTeodoro-21}.  The sample galaxies are selected from the publicly available HI surveys, including THINGS \citep[The HI nearby Galaxy Survey;][]{Walter-08}, HALOGAS \citep[the Hydrogen Accretion in LOcal GAlaxieS survey;][]{Heald-11}, LVHIS \citep[the Local Volume HI Survey; ][]{Koribalski-18}, and etc \citep{vanderHulst-01, Chung-09, Serra-12, Lutz-17}.  
These galaxies were selected mainly to have circular velocities greater than 100 km s$^{-1}$, an HI column-density sensitivity better than $\sim 5\times 10^{19}$ cm$^{-2}$, and to have inclinations INC between 30$^{\circ}$ and 80$^{\circ}$.  

Galaxies were examined visually by both of the authors independently in a blind way, i.e. all velocity fields were viewed in a random order and the distortions of the iso-velocity contours were classified to have an ``S", ``Z" or ``ambiguous" pattern.   This classification was repeated using a second random ordering of the optical images of the galaxies in order to determine the twist of the spiral arms.  For the spiral arms, the classification was generally unambiguous.  Our two independent classifications of all galaxies were then compared, giving also an indication of the reliability of the classifications.

For our primary test, the 28 (50\%) of the galaxies which showed clearly classifiable distortions of their velocity fields, and also had an unambiguous sense of their spiral arm pattern, were used. Of these, 13 galaxies showed the same sense of the spiral arms and the kinematic distortion, while 15 galaxies showed opposite senses, indicating that these are uncorrelated.  

With the proviso that spiral arms are assumed to be predominantly trailing, this test clearly suggests that the {\it dominant} effect in producing distortions of the velocity field is likely to be the presence of warps in the HI disks rather than the radial inflows whose signatures we had hoped to see.   Neither adding less easily classifiable galaxies, nor restricting the set to only the most beautiful examples, changed this conclusion.

\section{The hiding of radial motion when modelling the gas kinematic fields with pure warp model}
\label{sec:4}

In Section \ref{sec:3}, we showed that the signatures of radial motion in the velocity fields resemble those of warped disks. Observationally, the HI velocity fields (or HI datacubes) are frequently modelled with warped disk models without considering the possible systematic radial inflow \citep[e.g.][]{Jozsa-07, de-Blok-08, Kamphuis-15, Oh-18}, and then the residual velocity field is believed to show the level of non-circular motion \citep[e.g.][]{Schmidt-16, DiTeodoro-21}.  However, as shown in Section \ref{sec:3}, any systematic radial motion will also systematically twist both the kinematic major and minor axes.  In this section, we therefore try to understand, in general terms, how well a pure warp model will match the kinematic fields of disks that contain both systematic radial inflow and a warp, and therefore whether (or to what extent) radial motions could thereby be ``hidden'' by ascribing the distortion of the kinematic major and minor axes to the warped geometry alone.

\subsection{Construction of mock velocity fields and choice of fits} \label{sec:4.1}

We approach this question in the following way. We first construct mock datacubes of HI 21 cm emission for galaxy disks with known input radial inflow fields and with a known warped geometry of the disk. We then model these mock datacubes with a {\it pure warped disk} in order to examine whether (or under which conditions) the pure warped disk model can match the composite velocity fields produced with significant inflow. %These cases would represent systems where inflow was ``hidden" within the warp.

We both construct the mock datacubes and model their kinematics by using the code {\tt $^{\rm 3D}$BAROLO}\footnote{https://editeodoro.github.io/Bbarolo} \citep{DiTeodoro-15, DiTeodoro-21}.  
The code can efficiently construct a 3-dimensional mock datacube of a gas disk based on a set of concentric rings with given column density, geometry and kinematics.  

For simplicity, we input a constant column density of HI with radius in constructing the mock datacubes \citep{Bigiel-12}.  We adopt the Equation \ref{eq:5} with $V_{\rm cir}$ = 220 km s$^{-1}$ and $R_{\rm t}=0.5h_{\rm R}$ for circular motion. We use the radial velocity predicted in MAD with $h_{\rm R}=4$ kpc and $\lambda-R$=0.0 from Figure \ref{fig:1}. This particular radial velocity field is adjusted by randomly multiplying by a factor ${\tt vr\_scale}$, which is uniformly distributed between 0 and 2.  

\begin{figure}
  \begin{center}
   \epsfig{figure=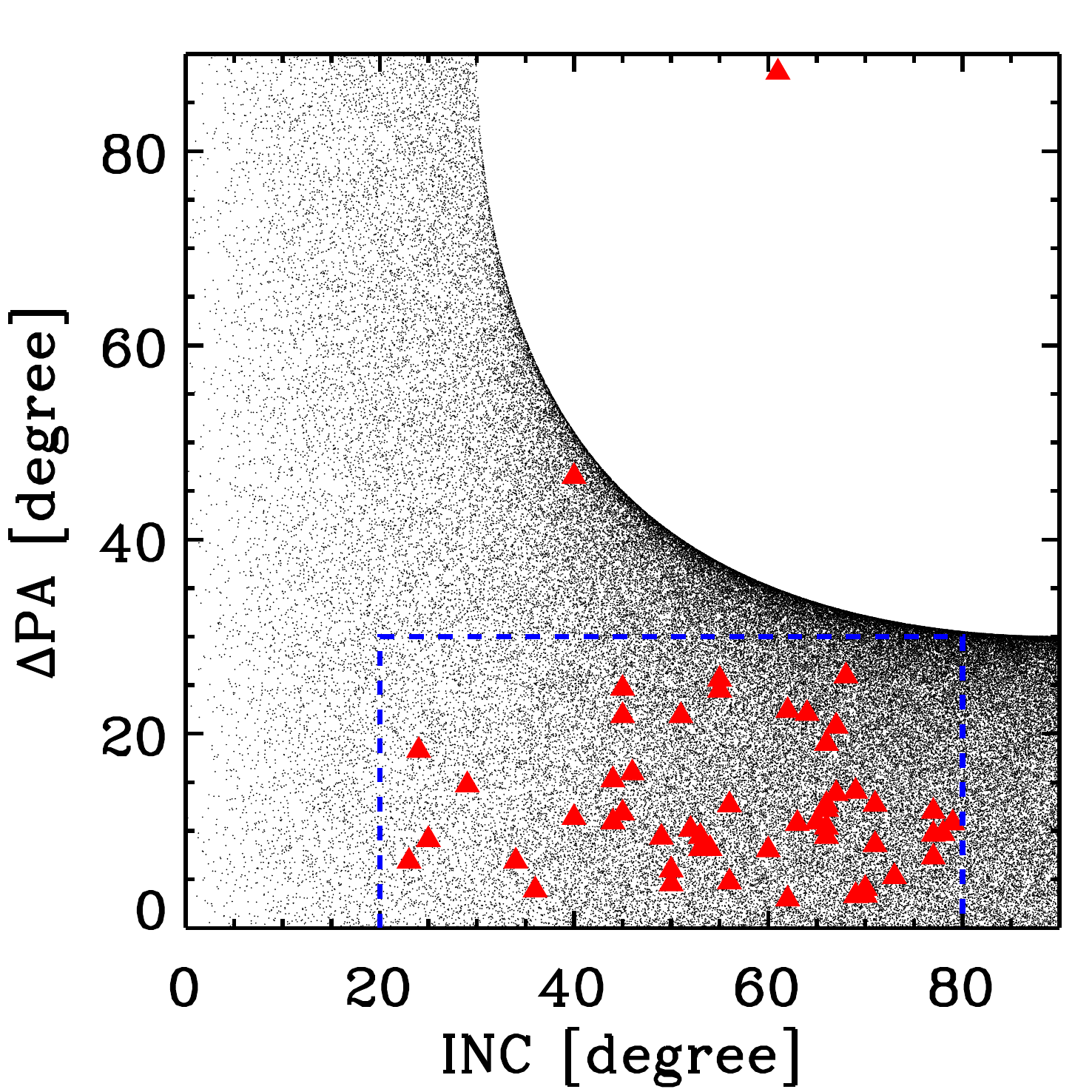,clip=true,width=0.38\textwidth}
  \end{center}
  \caption{ The $\Delta$PA and INC for the observed galaxy sample of \cite{DiTeodoro-21}.  The blue box shows the parameter space of our mock galaxies. The black small dots show the distribution of ``observed'' $\Delta$PA and INC for one individual system with an intrinsic $\Delta$PA of 30 degree but ``observed'' from all different directions.  }
  \label{fig:12.1}
\end{figure}

In order to input more realistic parameters, we first explore the radial changes of position angle (PA) and inclination (INC) for galaxies in the sample of \cite{DiTeodoro-21}. These are shown in Figure \ref{fig:12.1}.   \cite{DiTeodoro-21} fit the kinematic PA as a function of radius and we extract this radial change of PA (denoted as $\Delta$PA) from their fits.  $\Delta$PA is defined as the maximal range of returned position angles across the full range of radius.  We note that the $\Delta$PA obtained in this way is therefore gives an upper limit to the effect of the warps.  The black small dots in Figure \ref{fig:12.1} show the distribution of the ``observed'' $\Delta$PA and INC for a single individual 3d-system with an intrinsic $\Delta$PA of 30 degree when it is ``observed'' from all different directions.  

%As shown, the $\Delta$PA and INC of our mock galaxies are largely overlapped with those of real observed galaxies used in observational studies. 
As shown, the maximum $\Delta$PA for the real galaxies is $\sim$30 degree with two outliers, M83 and NGC 2685.  NGC 2685 is a polar ring galaxy \citep[e.g.][]{Jozsa-09}, and M83 shows outer disc streamers, an asymmetric tidal arm and a thoroughly twisted velocity field in HI map, which may be due to the accretion of dwarf galaxies in the past \citep{Koribalski-18}. 

We then construct a large number of mock data cubes according to Figure \ref{fig:12.1}.  For all, we input a constant position angle (PA, arbitrarily but unimportantly fixed to 90 degrees) for the mock galaxies within the radius of 5$h_{\rm R}$ (corresponding to the size of stellar disk), but then enable the PA to linearly increase or decrease beyond this radius.  The change of PA at 8$h_{\rm R}$, denoted as $\Delta$PA as above, with respect to 90 degree is set to be uniformly distributed from 0 to 30 degree across the set of mock cubes. Similarly, we input a constant INC for the mock galaxies within 5$h_{\rm R}$, and let it linearly increase or decrease
%\footnote{Here we only consider the case that INC decreases with radius, because we want to reduce the cases that the motions at different radii are overlapped on some regions of the LOS velocity fields. } 
beyond this radius. The INC of the mock galaxies is set to be uniformly distributed from 20 to 80 degrees within the radius of 5$h_{\rm R}$, and the change of INC at 8$h_{\rm R}$ with respect to the inner disk is set to be uniformly distributed from $-$10 to +10 degrees.  The blue box in the left panel of Figure \ref{fig:12.1} shows the parameter space of our mock galaxies.

Based on the above settings, we construct 200 mock datacubes. However, given the handedness discussed earlier, it is clear that the effect of the radial inflow can either be to enhance, or to suppress, the effect of the warp, depending on the sense of the disk rotation as seen by the observer.  We therefore construct two versions of each mock datacube: one where the two effects are in the same sense, i.e. the radial flow increases the kinematic distortion from the warp, and one where the effect of the inflow reduces the distortion from the warp. This doubles the number of mock galaxies. 

After constructing these mock datacubes, we then fit each of them with a model that contains only a warped disk (i.e. without any radial flow of gas) by adopting the {\tt 3DFIT} task under the code {\tt $^{\rm 3D}$BAROLO}, which is recommended to use \citep{DiTeodoro-21}.  This software fits 3-dimensional tilted-ring models to HI datacubes, resulting in the circular velocity, velocity dispersion, PA and INC for a given set of radial bins.  It should be noted that while this code can simultaneously fit for the radial motion, we do not enable this option for the purpose of our experiment at this stage.  

Although we use the {\tt $^{\rm 3D}$BAROLO} code of \cite{DiTeodoro-21}, it is important to note that we do not use the same procedure as in \cite{DiTeodoro-21}, because of our different goals. In the present work, we simply want to fit pure-warp models to the mock data cubes of composite warp+inflow models, and we therefore treat all pixels equally (without additional weights), in order to obtain the overall smallest residuals. This matches what has usually been done in the literature \citep[e.g.][]{de-Blok-08, Kamphuis-15}.  However, following the suggestion of the referee, 
%when extracting the radial velocities based on HI datacubes, 
we also present for comparison an additional version of the results that is obtained when we adopt a similar weighting scheme of pixels to that of \cite{DiTeodoro-21} (see details in Section \ref{sec:4.3}). 

%One of the key differences is the weighting of the pixels in the fits.  In an effort to extract the radial motion, \cite{DiTeodoro-21} employed a multi-step procedure and gave much higher weights to the pixels around the major axis in their first and second fits, and then gave higher weight to the pixels around the minor axis in the third fit.   In the present work, where we simply want to fit a pure-warp model to the mock data cubes of composite warp+inflow models, we always treat all pixels equally (without additional weights), in order to obtain the overall smallest residuals. This matches what is usually done in the literature \citep[e.g.][]{de-Blok-08, Kamphuis-15}.  Our tests and the results of \cite{DiTeodoro-21} should therefore not be directly compared.

\subsection{Nine galaxies for illustration} \label{sec:4.2}

\begin{figure*}
  \begin{center}
    \epsfig{figure=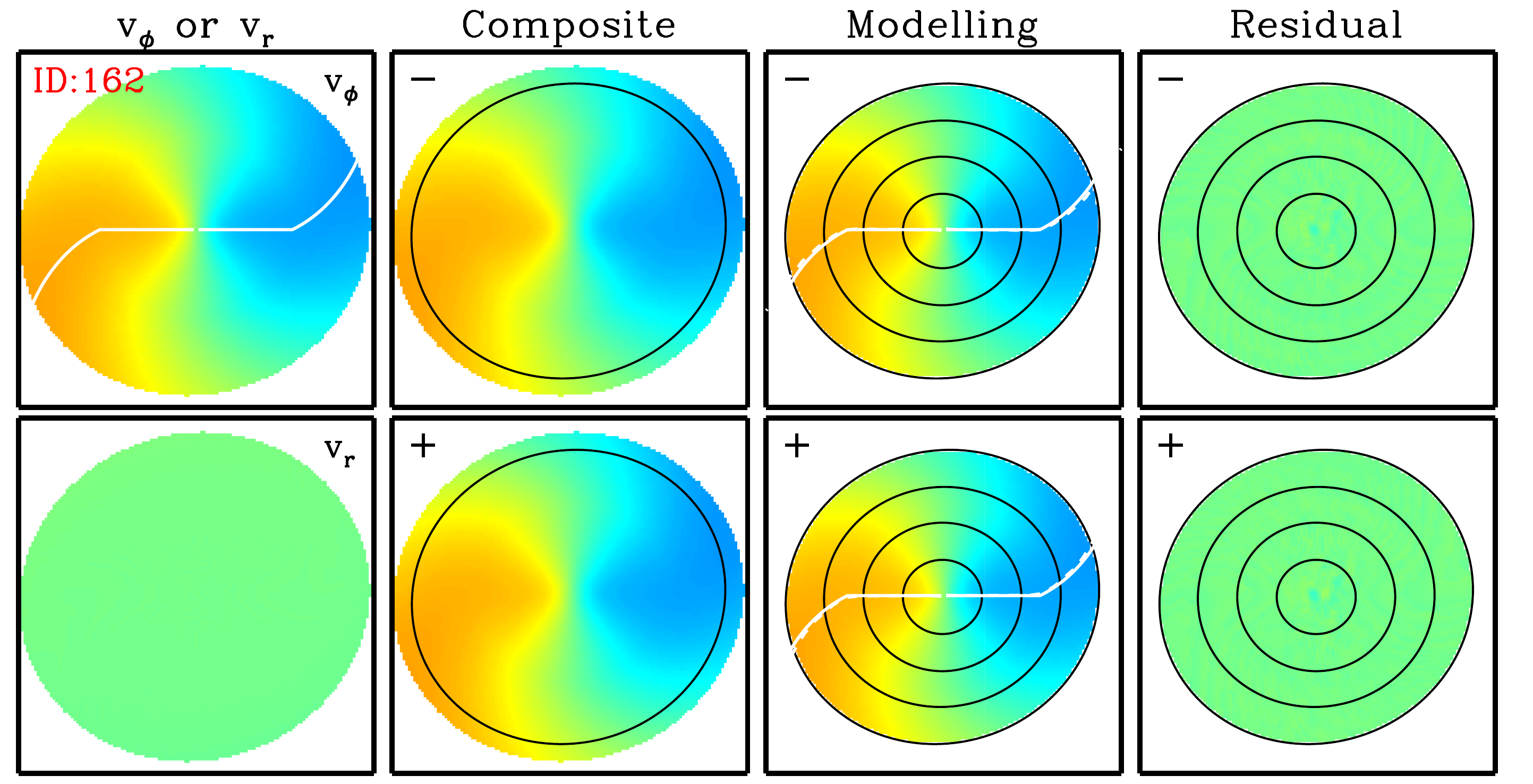,clip=true,width=0.76\textwidth}
    \epsfig{figure=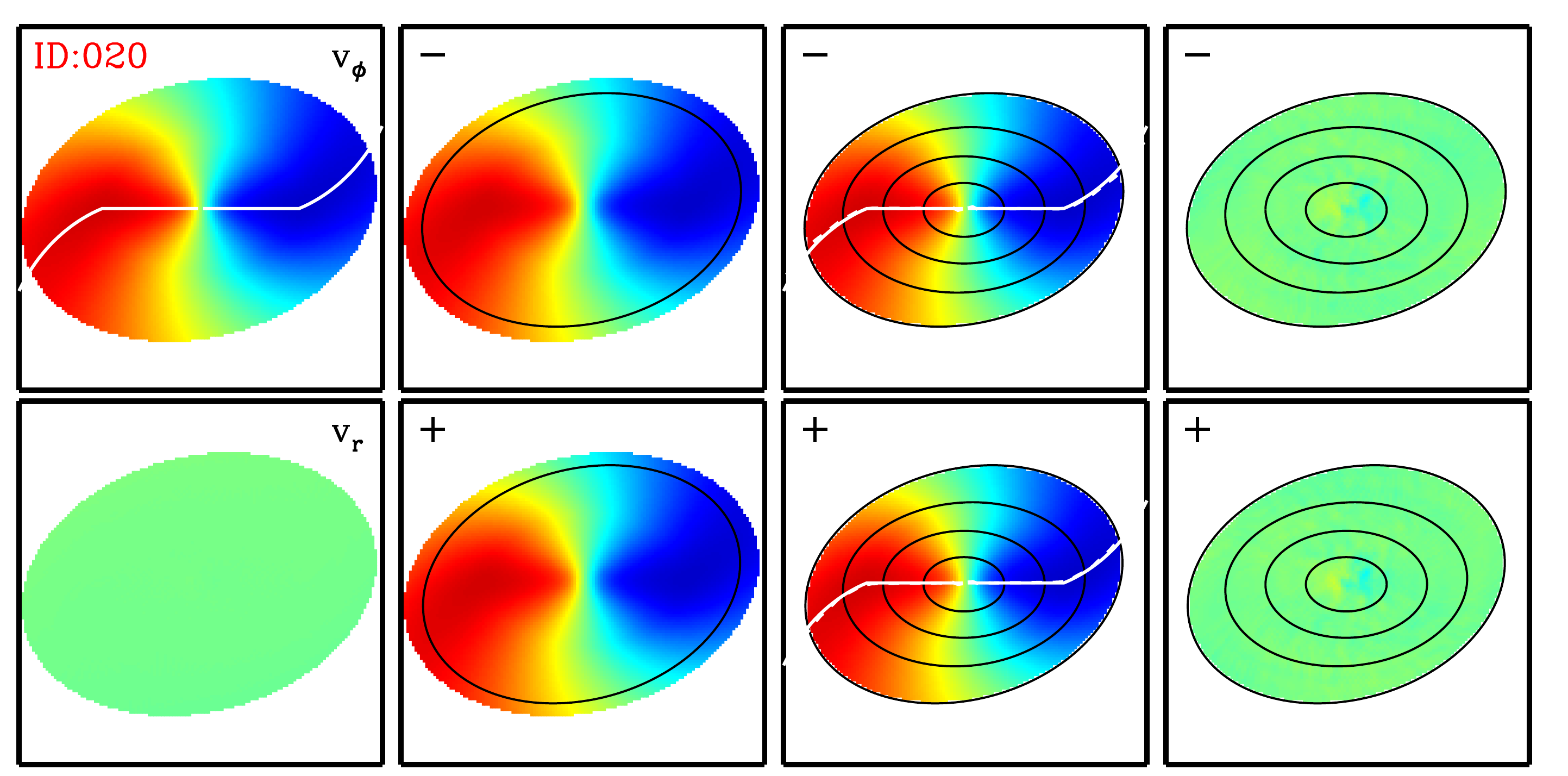,clip=true,width=0.76\textwidth}
    \epsfig{figure=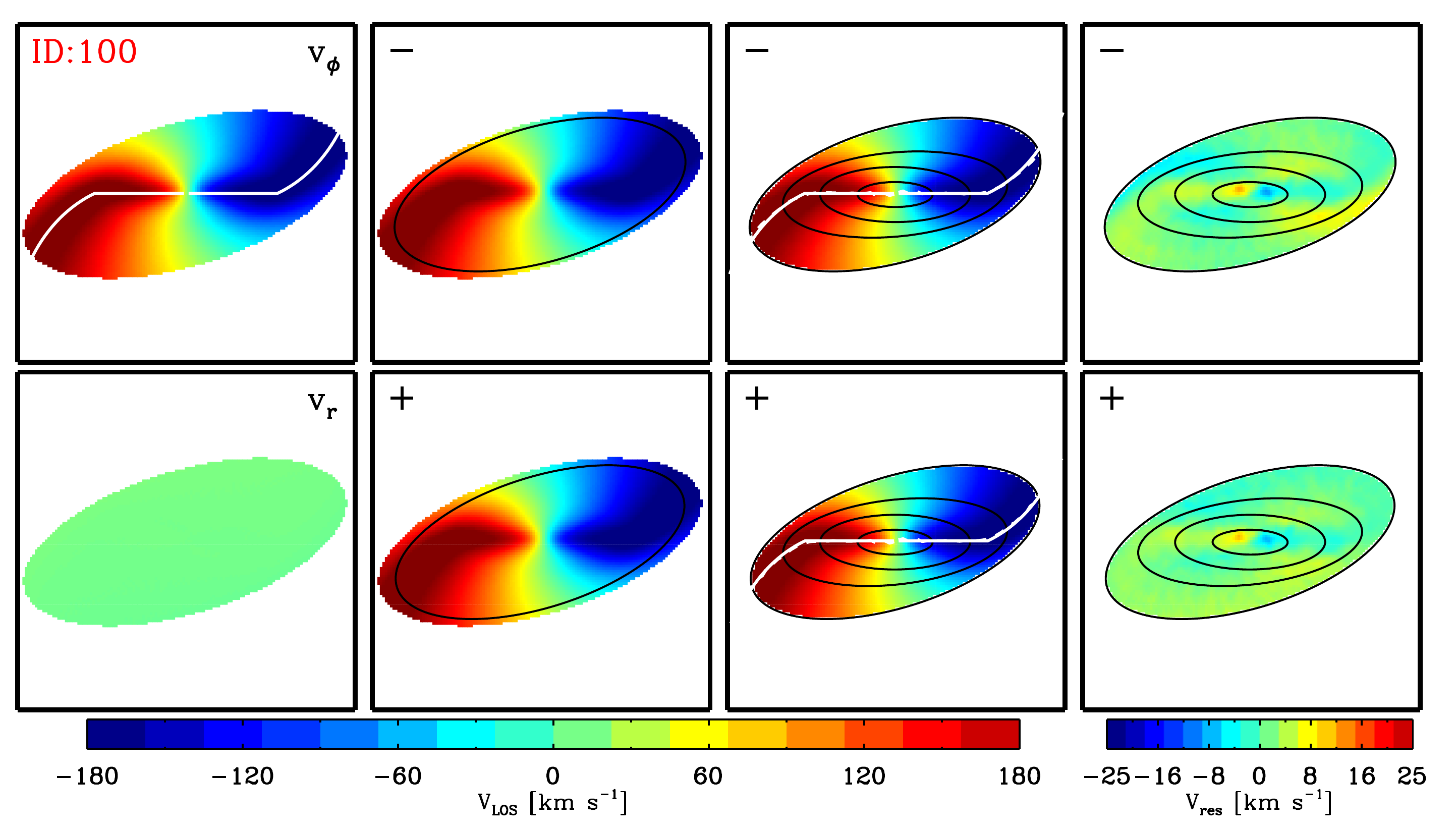,clip=true,width=0.76\textwidth}
  \end{center}
  \caption{The modeling of the velocity fields with 3D fitting for three selected three mock galaxies with low radial velocities (i.e. low ${\tt vr\_scale}$).  For each galaxy, we show the two cases in which the inflow reduces the distortion of the dominant warp (marked with ``$-$'') and in which it augments it (marked with ``$+$'').  For each row from left to right, the panels show the velocity field of circular motion (and the radial motion immediately below it), the composite velocity field in the two warp-enhancing and warp-suppressing cases, the fitted velocity field with pure warped disk model for both cases, and the residual map, respectively. The black ellipticals in the second column of panels from the left indicate the fitted regions, corresponds to the radius of 8$h_{\rm R}$ for the modeled field.
  In the two rightmost panels,  the four black ellipticals indicate the radii of 2$h_{\rm R}$, 4$h_{\rm R}$, 6$h_{\rm R}$ and 8$h_{\rm R}$ for the modeled velocity field. 
  In the velocity field of $v_{\phi}$, the white solid line shows the input geometric major axis of the corresponding mock galaxy, the same as the white solid line in the modeled velocity field, while the recovered major axis is shown in the white dashed line.  
  }
  \label{fig:4}
\end{figure*}

\begin{figure*}
  \begin{center}
    \epsfig{figure=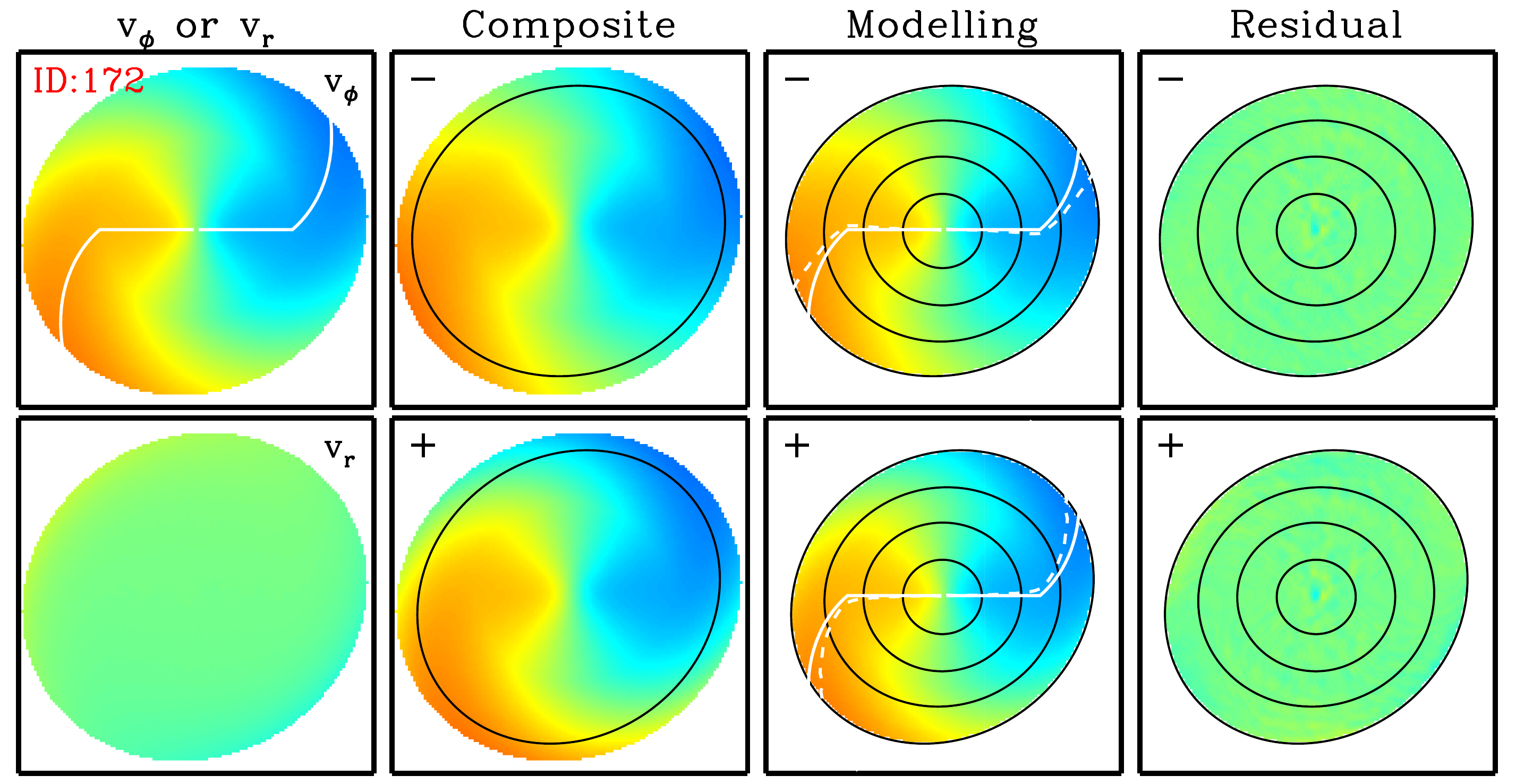,clip=true,width=0.76\textwidth}
    \epsfig{figure=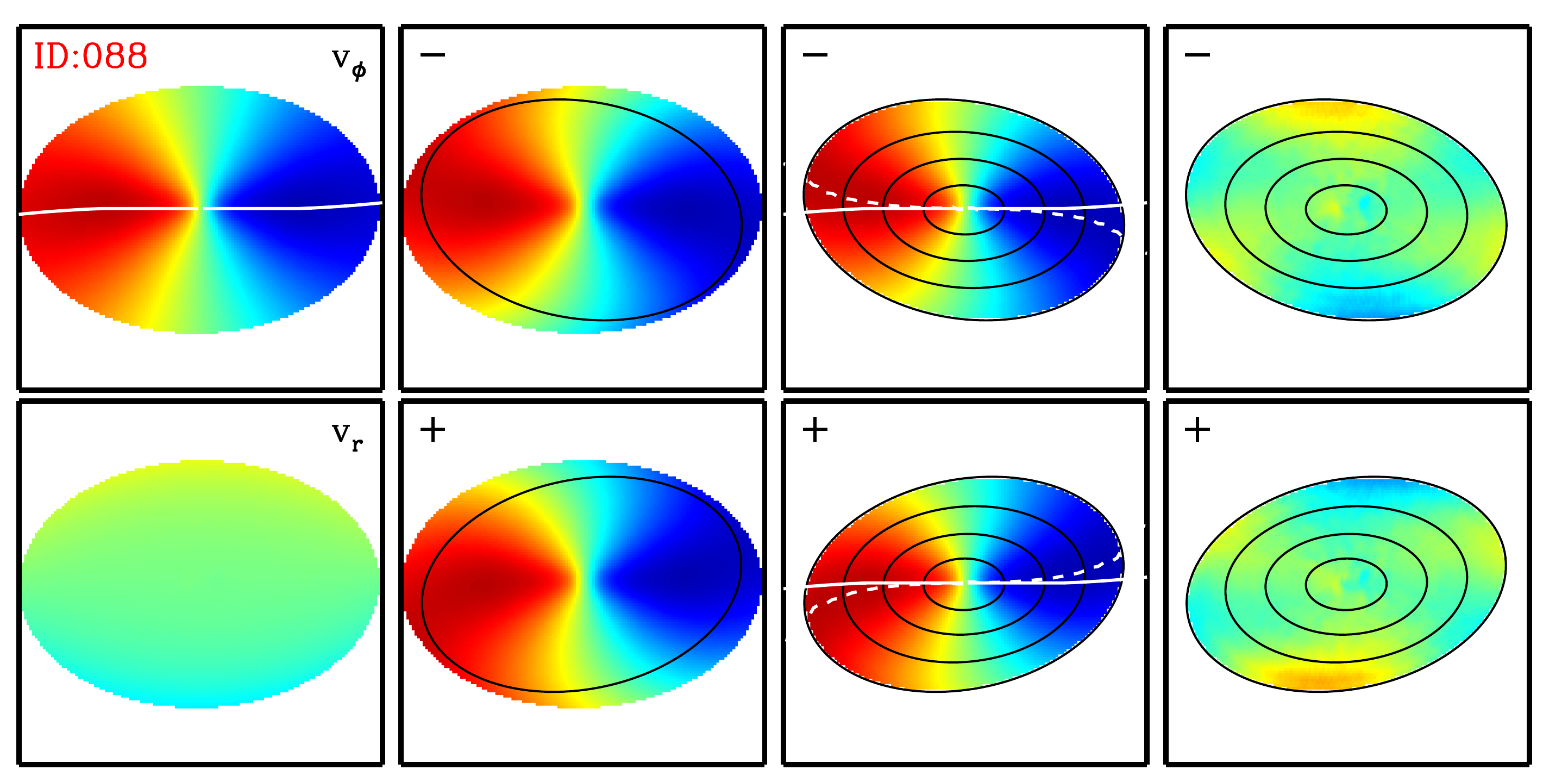,clip=true,width=0.76\textwidth}
    \epsfig{figure=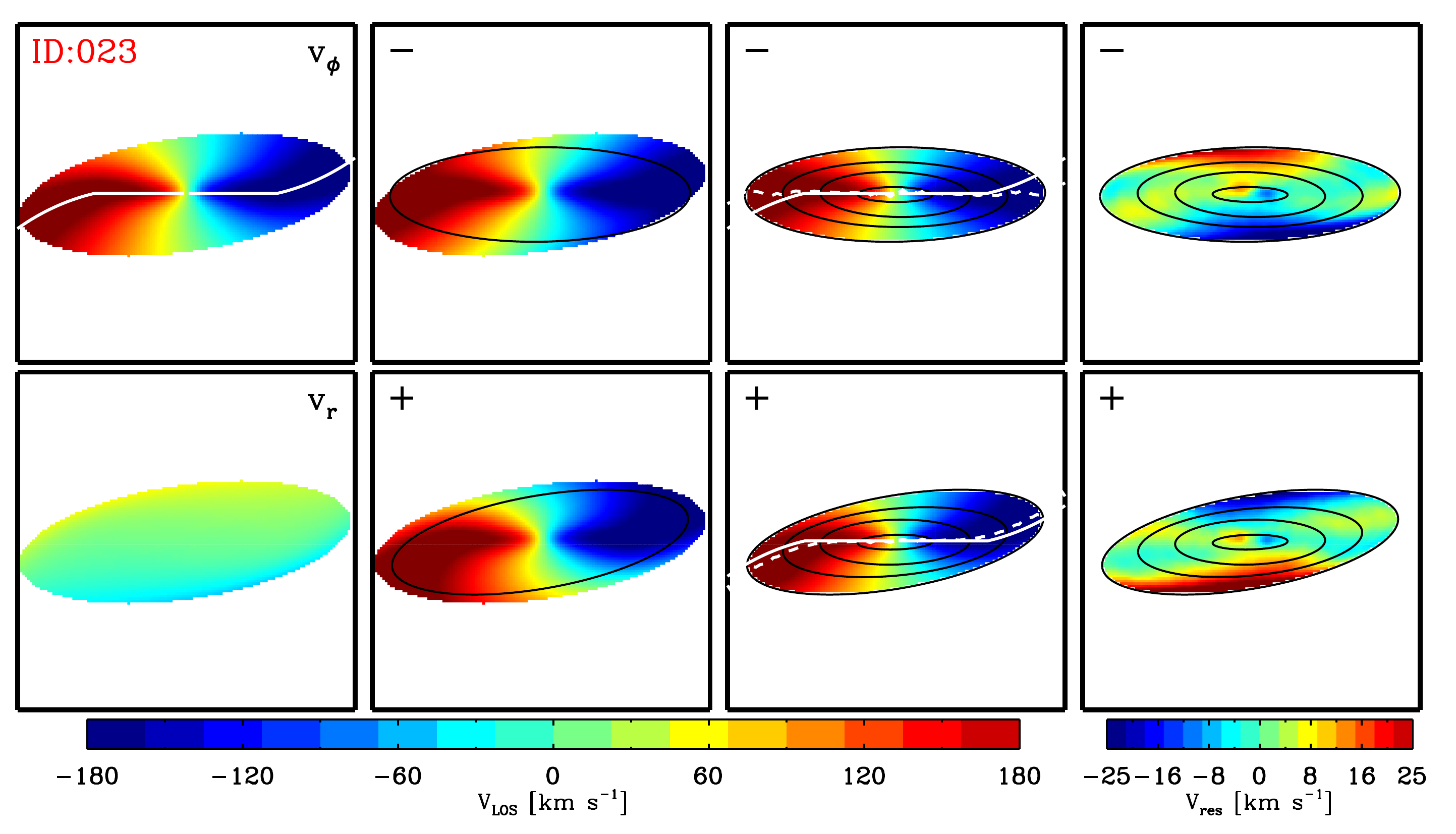,clip=true,width=0.76\textwidth}
  \end{center}
  \caption{ The same as Figure \ref{fig:4} but for three galaxies of intermediate ${\tt vr\_scale}$. }
  \label{fig:5}
\end{figure*}

\begin{figure*}
  \begin{center}
    \epsfig{figure=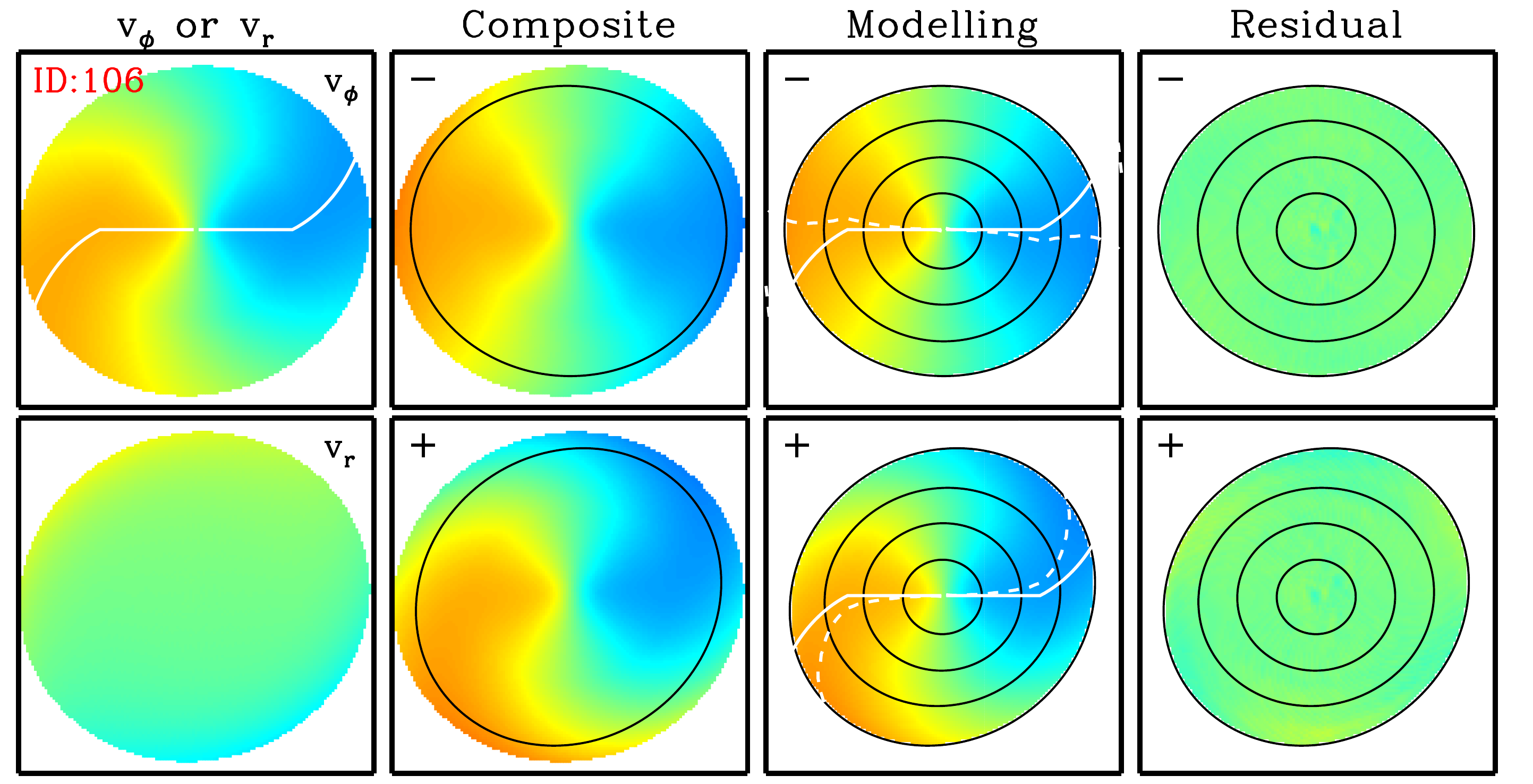,clip=true,width=0.76\textwidth}
    \epsfig{figure=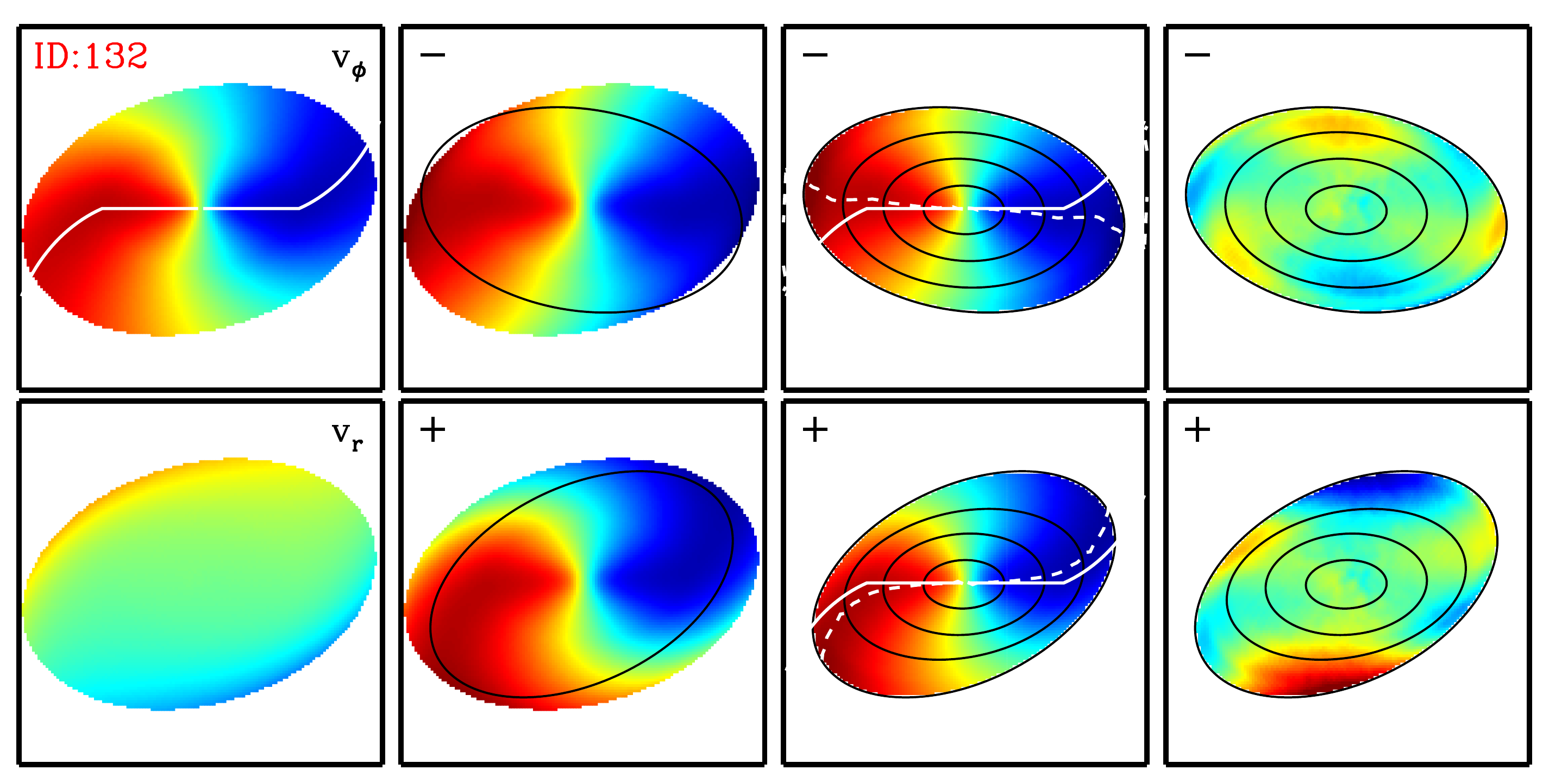,clip=true,width=0.76\textwidth}
    \epsfig{figure=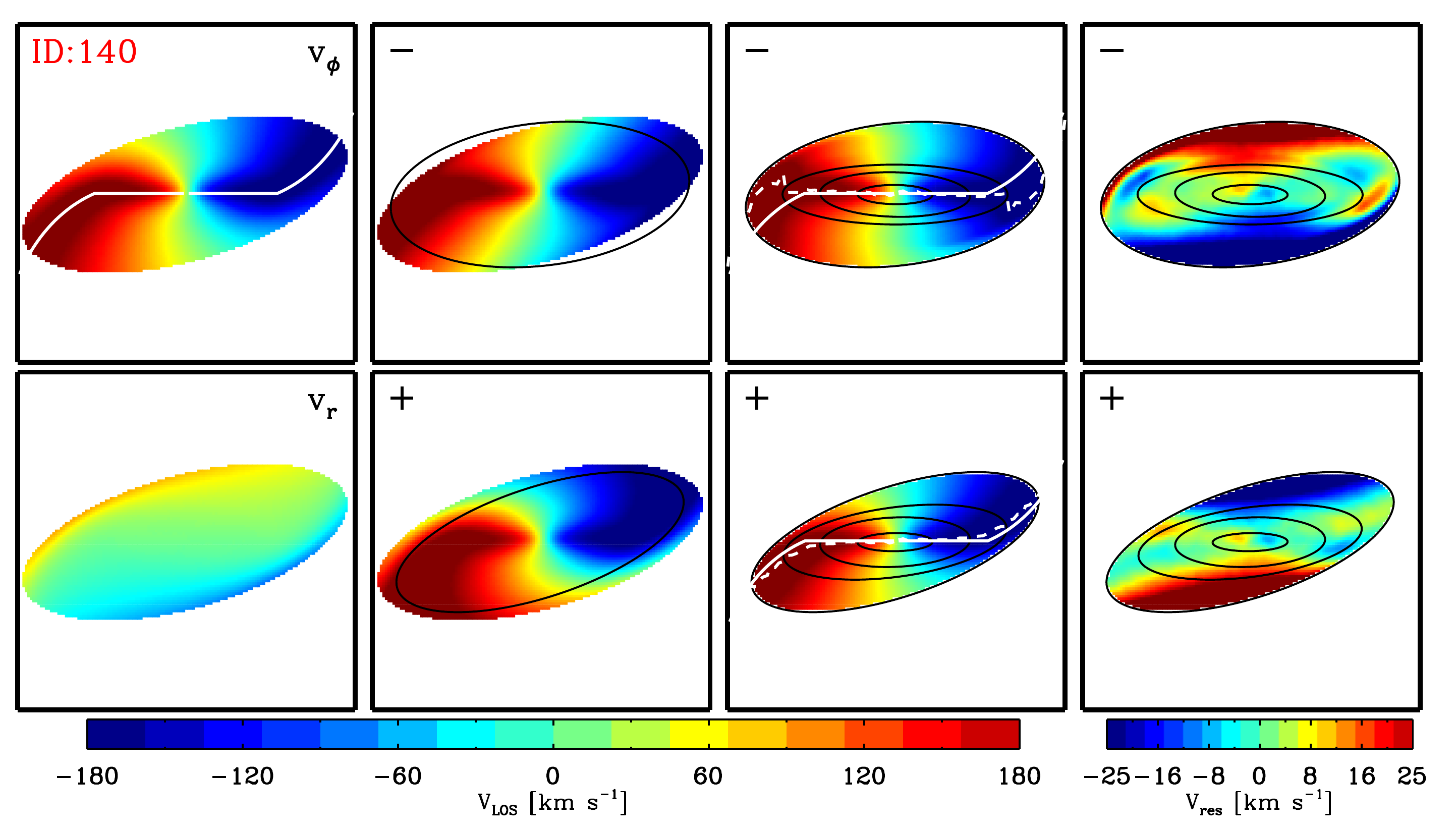,clip=true,width=0.76\textwidth}
  \end{center}
  \caption{ The same as Figure \ref{fig:4} but for the selected three galaxies with high ${\tt vr\_scale}$. }
  \label{fig:6}
\end{figure*}

\begin{figure*}
  \begin{center}
    \epsfig{figure=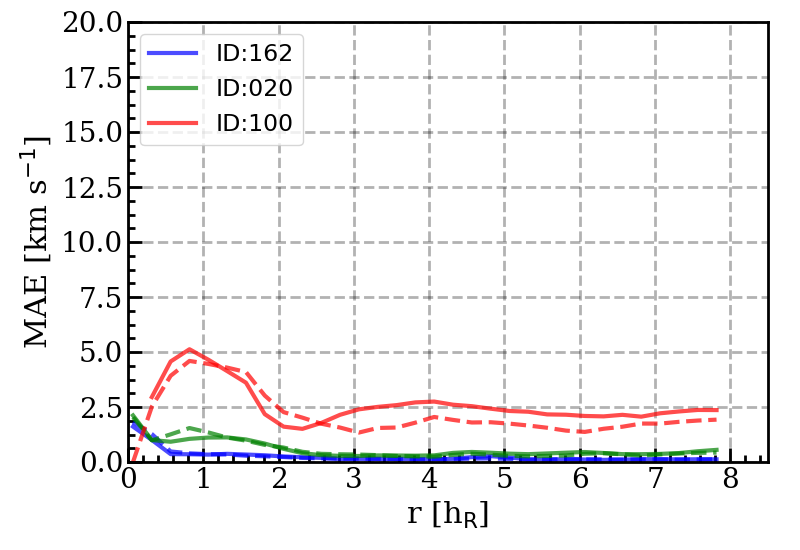,clip=true,width=0.33\textwidth}
    \epsfig{figure=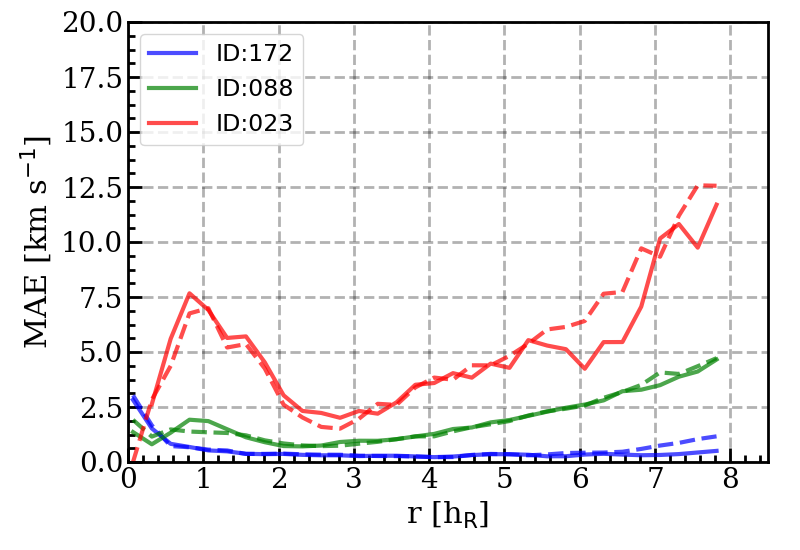,clip=true,width=0.33\textwidth}
    \epsfig{figure=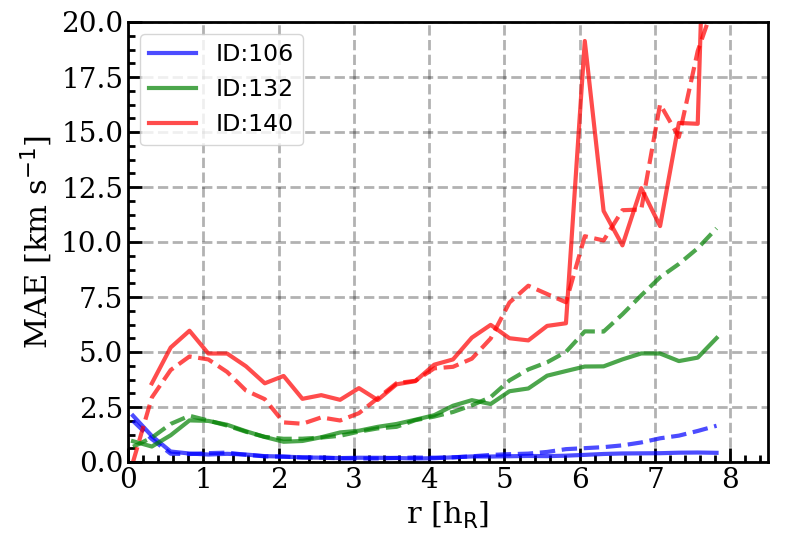,clip=true,width=0.33\textwidth}
  \end{center}
  \caption{ The MAE as a function of radius for the nine selected mock galaxies. The panels from left to right show galaxies with increasing ${\tt vr\_scale}$. In each panel, the galaxies are then shown with increasing INC (roughly from 20-70 degree) from the blue, green and red.  For each galaxy, the solid line shows the case where the flow suppresses the effect of the warp, and the dash line shows the case in which it enhances the warp. }
  \label{fig:7}
\end{figure*}

\begin{figure*}
  \begin{center}
    \epsfig{figure=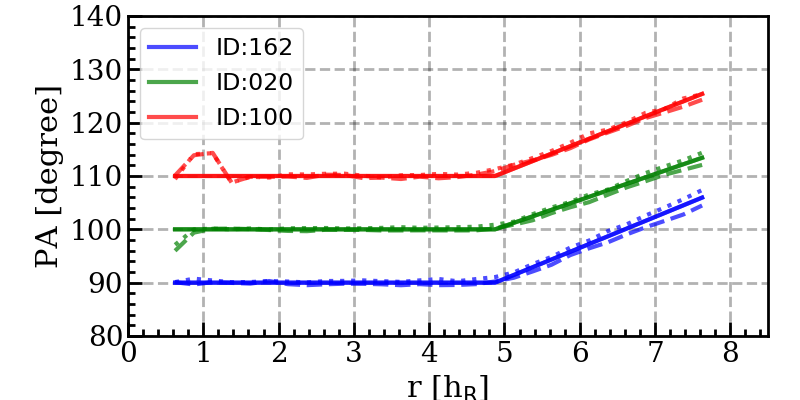,clip=true,width=0.33\textwidth}
    \epsfig{figure=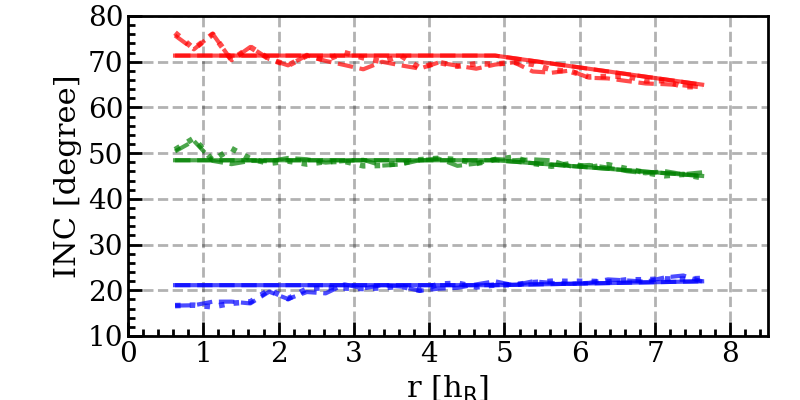,clip=true,width=0.33\textwidth}
    \epsfig{figure=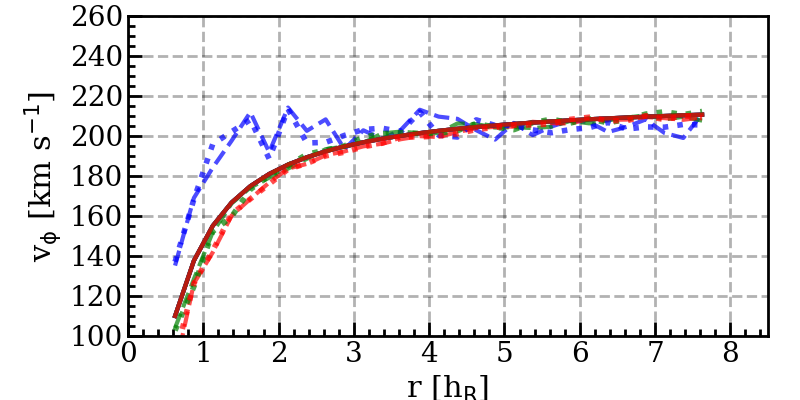,clip=true,width=0.33\textwidth}
    
    \epsfig{figure=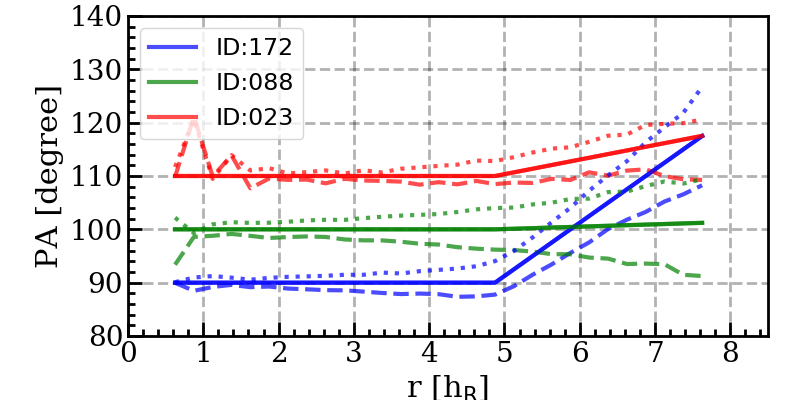,clip=true,width=0.33\textwidth}
    \epsfig{figure=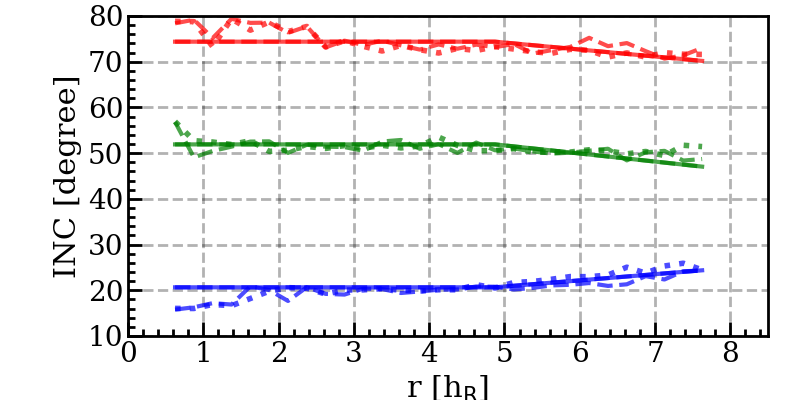,clip=true,width=0.33\textwidth}
    \epsfig{figure=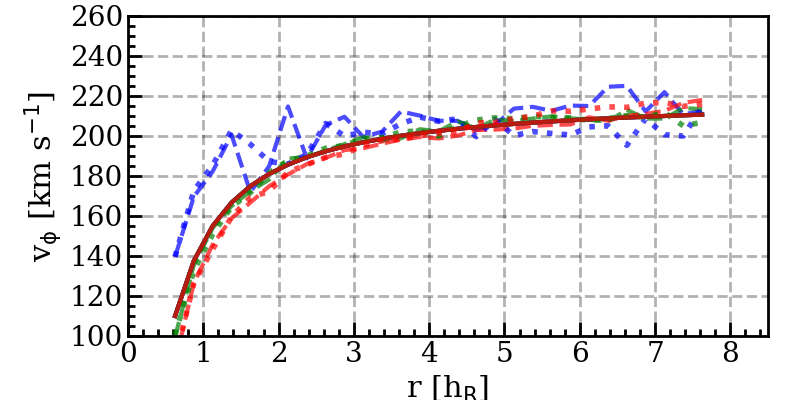,clip=true,width=0.33\textwidth}

    \epsfig{figure=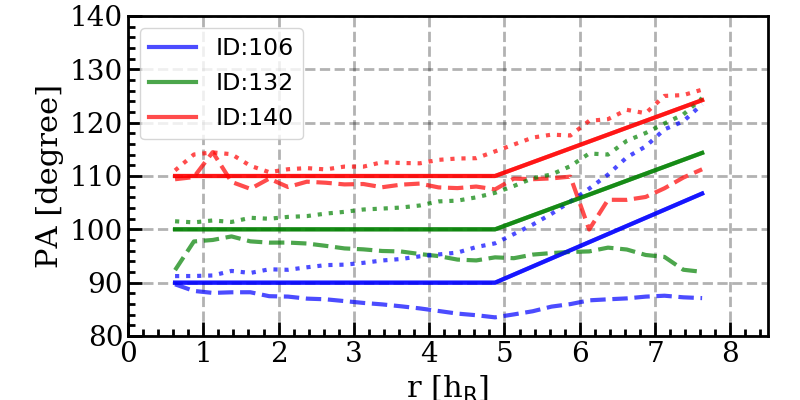,clip=true,width=0.33\textwidth}
    \epsfig{figure=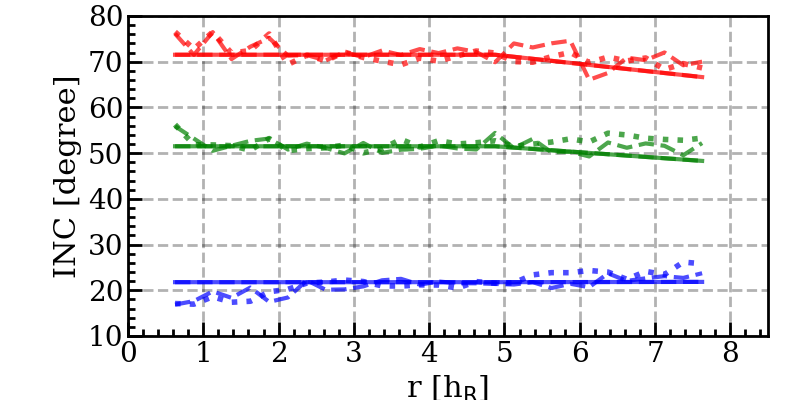,clip=true,width=0.33\textwidth}
    \epsfig{figure=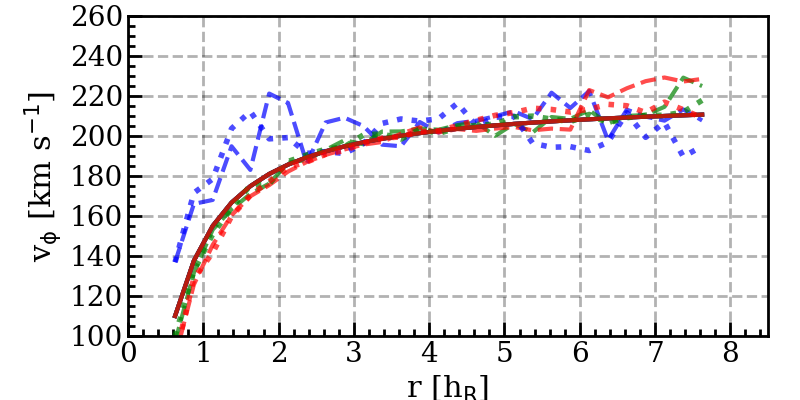,clip=true,width=0.33\textwidth}

  \end{center}
  \caption{The recovered PA, INC and $v_{\phi}$ (left to right in each row) as a function of radius for the nine selected mock galaxies when their composite velocity fields are fitted with pure warped disk models.  The IDs of the mock galaxies are indicated in the left column of panels only.  The solid lines show the input parameters, while the dashed and dotted lines show the fitted parameters when the flows suppress and enhance the warps, respectively.  From the top to the bottom the galaxies are shown with increasing ${\tt vr\_scale}$.  Within each row, the galaxies are shown with increasing INC (roughly from 20-70 degree) from blue, green to red, as in Figure \ref{fig:7}. In the left column of panels, the recovered PA profiles are shown with overall vertical shifts of 10 and 20 degrees for the galaxies with intermediate and high INC.  }

  \label{fig:8}
\end{figure*}

%\begin{figure*}
%  \begin{center}
%    \epsfig{figure=./fig/kinematic/mock/profile/res_vr_prof_lvr.png,clip=true,width=0.33\textwidth}
%    \epsfig{figure=./fig/kinematic/mock/profile/res_vr_prof_mvr.png,clip=true,width=0.33\textwidth}
%    \epsfig{figure=./fig/kinematic/mock/profile/res_vr_prof_hvr.png,clip=true,width=0.33\textwidth}
%  \end{center}
%  \caption{ The recovered radial velocities, normalized by the corresponding input radial velocity at  $\sim$8$h_{\rm R}$, as a function of radius for the nine selected mock galaxies. The panels from left to right show galaxies with increasing ${\tt vr\_scale}$. In each panel, the galaxies are then shown with increasing INC (roughly from 20-70 degree) from blue, green to red. For comparison, we show the input radial velocity as the black solid line in each panel.  
%  For each galaxy, the dashed line shows the case where the flow suppresses the effect of the warp, and the dotted line shows the case where it enhances it. }
%  \label{fig:7.1}
%\end{figure*}

% 0. Selection of nine galaxies, spanning a large range of parameters
In this subsection, we show the fitting results and the recovered parameters in the above test obtained with uniform weighting of the pixels.  For illustration, we show the fitting results for a representative sample of nine galaxies, which span a large range of both ${\tt vr\_scale}$ and INC. The selected nine galaxies can be separated into three groups according to the values of ${\tt vr\_scale}$: ${\tt vr\_scale}\sim$0.1, ${\tt vr\_scale}\sim$1 and ${\tt vr\_scale}\sim$1.8.  For orientation, ${\tt vr\_scale}\sim$1 (or ${\tt vr\_scale}\sim$1.8) corresponds to $v_r\sim$45 km s$^{-1}$ (or 80 km s$^{-1}$) at 8$h_{\rm R}$ (see the blue solid line in Figure \ref{fig:1}). 
For each group, three mock galaxies are selected to have very different INC (within 4$h_{\rm R}$). As will be shown later, the INC is the most important parameter in determining the goodness of the pure-warp fits.  %The $\Delta$PA, INC and ${\tt vr\_scale}$ for the selected nine galaxies are indicated later in Figure \ref{fig:10}. 

\subsubsection{ The fitting of the 3d datecubes with pure-warp models}  \label{sec:4.2.1}
%1. The velocity fields and fittings
Figure \ref{fig:4} shows the fitted velocity fields for three selected galaxies with low ${\tt vr\_scale}$.  For each galaxy, we show the two cases in which the radial inflow has either reduced ($-$) or enhanced ($+$) the distortions caused by the warp, with all other parameters the same.   For each of the three galaxies, the panels show from left to right the velocity field of circular motion (above) or radial motion (below), and then for both reduced and enhanced cases, the composite velocity field, the velocity field fitted with a pure warped disk model, and the residual map, respectively.  These three galaxies are displayed from top to bottom with increasing INC. 
In the velocity field of $v_{\phi}$, the white solid line shows the input geometric major axis of the mock galaxy, the same as the white solid line in the modeled velocity field, where the recovered major axis is also indicated by the white dashed line. We note that the 2-dimensional velocity fields are shown for illustration, while the fits are always performed in 3-dimensions. 

As shown in Figure \ref{fig:4}, the modeled velocity fields using the pure-warp model are overall in very good agreement with the constructed composite inflow+warp velocity fields for these three galaxies. The fitted kinematic major axes closely follow the actual ones, and the residuals are extremely small (only a few km s$^{-1}$).  This is not surprising, because these three galaxies were constructed with small radial motions. In other words, Figure \ref{fig:4} primarily indicates the robustness of the fitting code we used.  

Figure \ref{fig:5} and Figure \ref{fig:6} are similar to Figure \ref{fig:4}, but are for additional mock galaxies with intermediate and high ${\tt vr\_scale}$, respectively.  As we increase the strength of the radial motion, the modeled kinematic major axes show clear deviations from the input geometric major axis, because of the extra contribution of the radial flow.  But the overall 2-d velocity fields are still very well recovered by the pure-warp model, especially for galaxies of low and intermediate INC.  In other words, the effects of the inputted radial motions are being treated without difficulty as a contribution from the warped disk in the fitting for low-to-intermediate inclined disks, with reasonably small residuals (we will quantify the residuals in later Section \ref{sec:4.2.2}).  This result is consistent with our theoretical expectation from Section \ref{sec:3} that radial motions cannot be easily separated from the effects of a warped disk based on kinematic features alone, at least for reasonably weakly inclined disks.  %If a warp-plus-inflow system is equally well fittable as a pure-warp system, then given the difficulty of independently knowing a priori the parameters of the warp, it is clear that radial flows can be effectively ``hidden" within the uncertain warps.

For highly inclined galaxies with significant radial inflows, there are significant bi-polar symmetric residuals near the minor axis when the composite velocity field is fit with a pure-warp model. %The maximum value of the residuals depend strongly on the INC of galaxies with increasing from 10\% or 20\% to as large as 70\% of input inflow velocity for the mock galaxies. 
Such residuals could be taken as a signature of radial motion in observations.
We have explored HI residual maps in the literature and find that some individual galaxies show a similar symmetric feature near to the minor axis of the residual velocity fields. A good example is the Circinus galaxy \citep{Kamphuis-15, Koribalski-18}. As can be seen in the figure 8 of \cite{Kamphuis-15}, the Circinus galaxy show clear symmetric residuals close to the minor axis with a range of different fitting codes.

However, one should realize that the recovered kinematic major axes still show significant deviations from the input geometric ones, which further suggests that even if the signatures of radial motion are found in the residual maps, the extraction of {\it accurate} estimates of the radial motion may still be challenging.  We will later extract the radial motions from the residual maps, in order to examine what fraction of the radial inflow has been hidden by fitting with the pure warp model. 

\subsubsection{The residuals and the recovered parameters}  \label{sec:4.2.2}

%1. The MAE of the fittings 
We introduce the mean absolute error (MAE) to quantify the residuals. This is defined as the mean value of the absolute deviations of the fitted pure-warp velocity field from the actual input warp+inflow composite velocity field at a particular radius.
Figure \ref{fig:7} shows the MAE as a function of radius for the nine selected galaxies shown in Figure \ref{fig:4}-\ref{fig:6}.  The galaxies are shown with increasing ${\tt vr\_scale}$ from left to right. In each panel, galaxies are then color-coded by inclination INC.  For each galaxy, the solid line shows the case where the inflow suppresses the effect of the warp and the dashed line shows the case where the inflow enhances it. 

As seen in Figure \ref{fig:7}, the MAE increases significantly with the inclination, INC. This is only partly due to the fact that the amplitude of all LOS velocities increase with INC. As we discussed later in Section \ref{sec:4.3}, the fits depend on the INC even with considering the above effect. This is in accord with the theoretical expectation in Section \ref{sec:3} that the effects of radial motion become increasingly different from those of the warped disk as INC increases.  

For galaxies with significant radial motion (see the middle and right panels of Figure \ref{fig:7}), the MAE increases with radius (or inflow velocity), and with ${\tt vr\_scale}$.  For the most inclined galaxies, the MAE is as large as 12 km s$^{-1}$ at 8$h_{\rm R}$ when the radial velocity reaches  $\sim$45 km s$^{-1}$ (see the mock galaxy of ID: 023), and the MAE is 20 km s$^{-1}$ at 8$h_{\rm R}$ when the radial velocity reaching $\sim$68 km s$^{-1}$ (see the mock galaxy of ID: 140).  
We can conclude that, at least for low-to-intermediate inclined disks, the radial motion can, to a very large extent, be hidden when fitting the composite HI datacubes with the pure warp disk model. 

%3. The recovered delta PA, INC and Vrot 
Force-fitting the mock datacubes with warp-only models also returns perturbed values of PA, INC and circular velocity as a function of radius. Figure \ref{fig:8} shows the comparisons of the returned values of these parameters with the input ones for the same nine galaxies.
The solid lines show the input parameters as a function of radius, while the dashed and dotted lines show the returned perturbed ones for the two cases in which the radial inflow suppresses and enhances the effect of the warps, respectively.  From the top to bottom row, the galaxies are shown with increasing ${\tt vr\_scale}$.  In each panel, the galaxies are shown with increasing INC from blue, green to red, as above.  

As can be seen in Figure \ref{fig:8}, the returned circular velocity and INC at different radii are generally not strongly perturbed by the force-fit of the pure-warp model. Weakly-inclined galaxies show a slightly larger deviation of the returned $v_{\phi}$ (blue lines) from the true input value. This increase is mostly due to the correction of velocities for the inclination of the galaxies.  However, the recovered PA (as a function of radius) shows significant deviations when the inflow velocity is significant. This again illustrates the point made earlier in Section \ref{sec:3} that the kinematic major axis is noticeably distorted from the geometric major axis in the presence of significant radial motion (they should be the same in a pure-warp model), even though radial motions have no contribution to the LOS velocity map along the geometric major axis.

% ---- adding the inflow velocity from the residuals ----
The MAE does not fully characterize the relevant features of the residual maps in Figure \ref{fig:4}-\ref{fig:6}, because of the existence of bi-polar symmetric residuals near the minor axis.  To account for this, we can extract the radial motion from the residual maps. 
By fixing the PA, INC and rotation velocity obtained from the fitting with pure-warps model (see Figure \ref{fig:8}), we then perform an additional fit for the radial velocity only with uniform weighting of the pixels. We would expect the recovered radial velocities to be significantly smaller than the real (input) ones for two reasons. First, the fitted warps already contain the effect of the radial motion, so the amplitude of the residual map is correspondingly less than the LOS contribution of the input radial velocity. Second, the radial velocities are mainly extracted from the fitted (kinematic) minor axis rather than the geometric minor axis.  

Again, we stress that the above method is similar to but different from the method of \cite{DiTeodoro-21}. One of the key differences is that we do not weight the individual pixels during the fitting, so as to obtain the smallest overall residuals.  For comparison, we also extract the radial motion from the datacubes following the same weighting scheme as in \cite{DiTeodoro-21}.  Specifically, we first fit the datacubes to obtain the PA, INC, circular velocity and velocity dispersion as a function of radius with pure-warp model by weighting the pixels with $cos^2(\theta)$, where $\theta$ is the angle to the kinematic major axis.  We then perform an additional fitting to obtain only the radial velocities by weighting the pixels with $sin^2(\theta)$, fixing all other parameters to the values previously determined.  We will show the recovered radial velocities with and without the weighting scheme in the next subsection. 

Even when using the same weighting scheme as in \cite{DiTeodoro-21},  our approach is still not exactly the same as the three-step fitting method in \cite{DiTeodoro-21}. 
In particular, we do not include the second-step fit of \cite{DiTeodoro-21}, which appears to produce little improvements in the cases examined.  \cite{DiTeodoro-21} excluded all pixels that lie within 20 degree from the kinematic minor axis in their first and second fittings, and excluded the pixels that lie within 20 degree from the kinematic major axis in their third fitting. We do not exclude any pixels in any of our fittings.  However, since the weighting scheme has already given these pixels \citep[excluded in the fittings of][]{DiTeodoro-21} very low weights, we do not expect that this would cause a very significant difference.  

%Figure \ref{fig:7.1} shows the extracted radial velocity, ${\tt Vr\_{out}(r)}$, as a function of radius for the nine galaxies. The radial velocities have been normalized by the corresponding input radial velocity at $\sim$8$h_{\rm R}$, ${\tt Vr\_{in}(8h_{\rm R})}$. For comparison, we show the input radial velocity in the black solid line.  The galaxies are displayed in the same order as in Figure \ref{fig:7}. 

%As shown, for the cases of small ${\tt vr\_scale}$ (the left panel), the recovered radial velocities are overall close to zero across the full range of radii, except for the mock galaxy with highest INC (ID:100). This galaxy shows very large radial variations in ${\tt Vr\_out(r)/Vr\_in(8h_{\rm R})}$, which is due to the fact that the input radial velocity is too small, and the residuals become significant with respect to the input radial velocity. For the more interesting cases of intermediate and high ${\tt vr\_scale}$, the resulting ${\tt Vr\_out(r)/Vr\_in(8h_{\rm R})}$ appears to strongly depend on INC, while showing no clear dependence on ${\tt vr\_scale}$.  For low-to-intermediate inclined galaxies (INC$<$50 degree), the recovered ${\tt Vr\_out(r)}$ are only $\sim$20\%, or less, of the real input velocities, indicating that most of the inflow velocity has been hidden within the warp.  The recovered ${\tt Vr\_out(r)}$ increases to $\sim$60\% for highly inclined galaxies (INC$>$70 degree).   This indicates a strong INC-dependence for the hiding of radial motions within warped disks. We will give more quantitative analysis of this in the next subsection. 

\subsection{The extracted radial motion from the mock galaxies} \label{sec:4.3}

\begin{figure*}
  \begin{center}
    \epsfig{figure=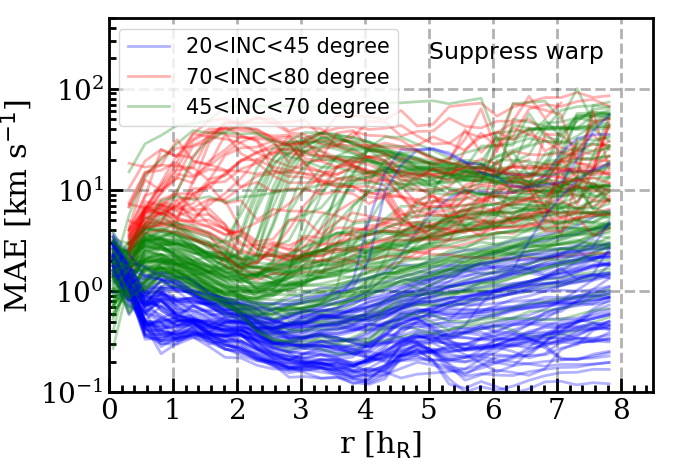,clip=true,width=0.33\textwidth}
    \epsfig{figure=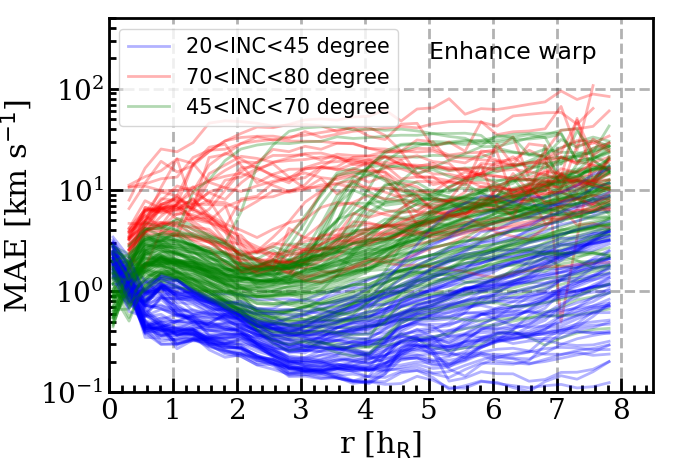,clip=true,width=0.33\textwidth}
    \epsfig{figure=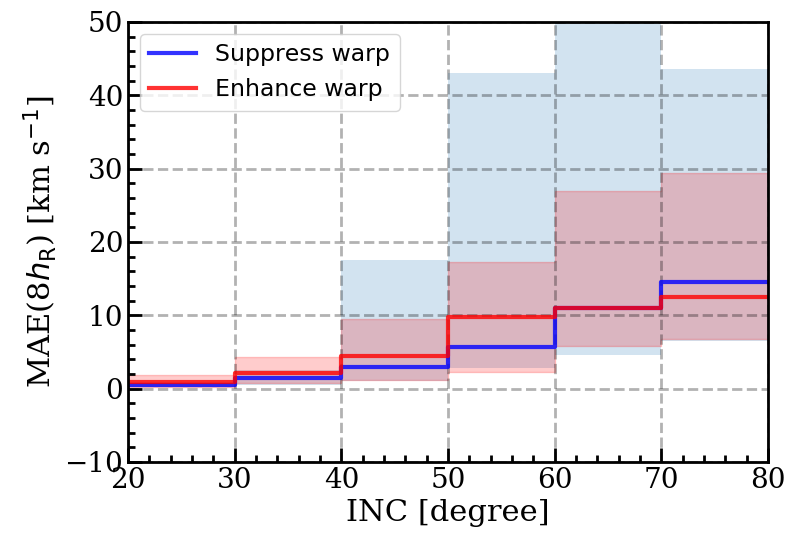,clip=true, width=0.33\textwidth}
  \end{center}
  \caption{  Left and middle panels: the MAE as a function of radius for all 200 mock galaxies.  The left panel is for the case of inflows suppressing the effect of the warps, and the middle panel is for the case of enhancing the effects of the warps. In the left and middle panels, the galaxies are separated into three subsamples according to INC, as denoted in the figure. Right panel: the median MAE(8$h_{\rm R}$) as a function of the input inclination, for the cases of both suppressing and enhancing warps.   In the right panel, the shaded regions show the 16-84\% percentiles of the mock galaxies. We bin the mock galaxies according to their INC with the bin width of 10 degree. For each bin, the median and percentiles are thus calculated based on the mock galaxies located in that bin. }
  \label{fig:9}
\end{figure*}

\begin{figure*}
  \begin{center}
    \epsfig{figure=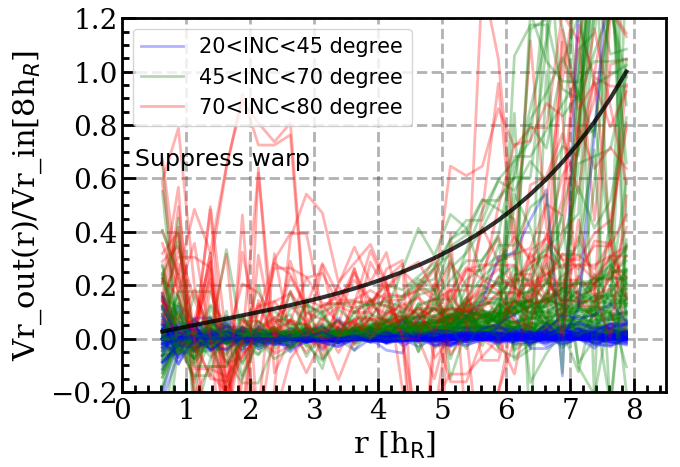,clip=true,width=0.33\textwidth}
    \epsfig{figure=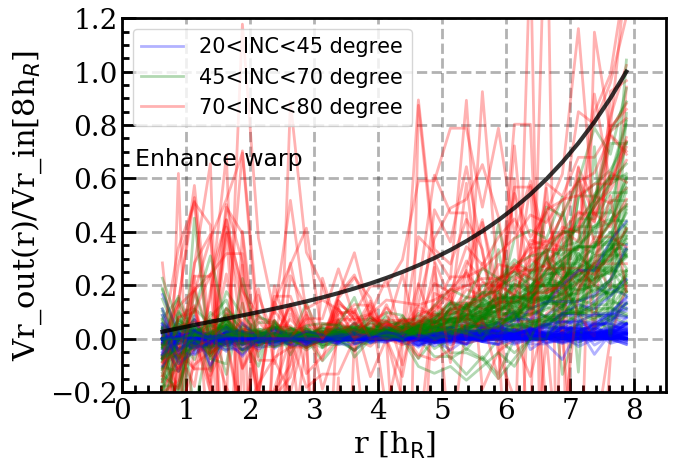,clip=true,width=0.33\textwidth}
    \epsfig{figure=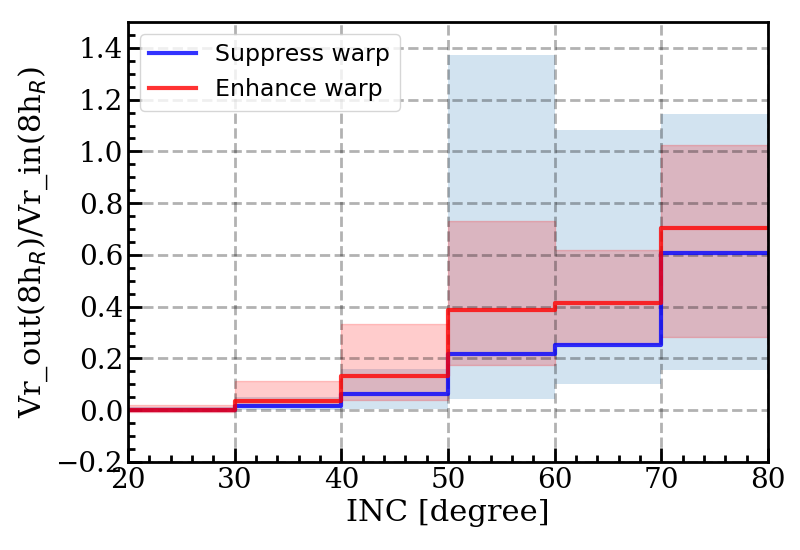,clip=true, width=0.33\textwidth}
  \end{center}
  \caption{  Left and middle panels: the extracted inflow velocity ${\tt Vr\_{out}(r)}$ obtained with uniform weighting of the pixels, normalized by the input $v_r$ at $\sim$8$h_R$,  as a function of radius for all 200 mock galaxies. Similar to Figure \ref{fig:9}, the left panel is for the case of inflows suppressing the effect of the warps, and the middle panel is for the case of them enhancing the effects of the warps. In the left and middle panels, galaxies are separated into three subsamples according to their INC.  Right panel: the median ${\tt Vr\_out(8h_{\rm R})/Vr\_in(8h_{\rm R})}$ as a function of inclination.  The shaded regions show the 16-84\% percentiles of the mock galaxies, calculated in a similar way as in Figure \ref{fig:9}.  } 
  \label{fig:10}
\end{figure*}

\begin{figure*}
  \begin{center}
    \epsfig{figure=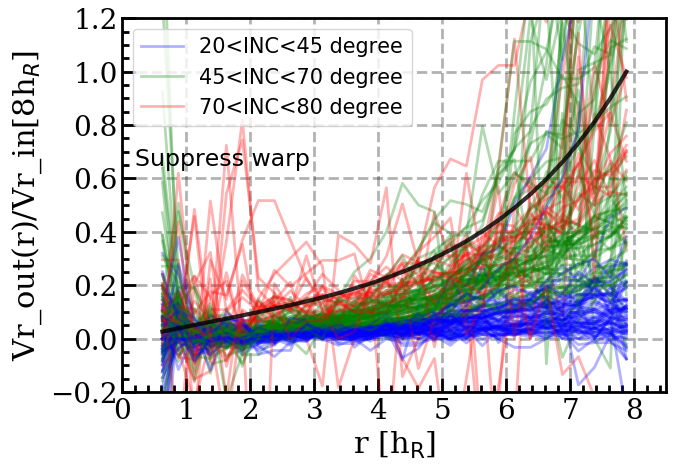,clip=true,width=0.33\textwidth}
    \epsfig{figure=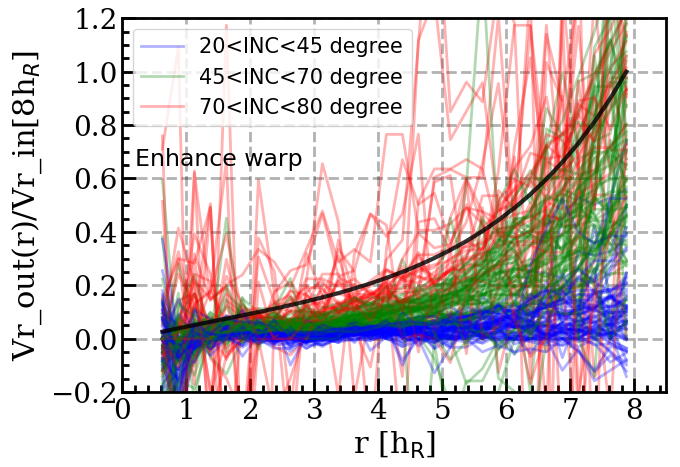,clip=true,width=0.33\textwidth}
    \epsfig{figure=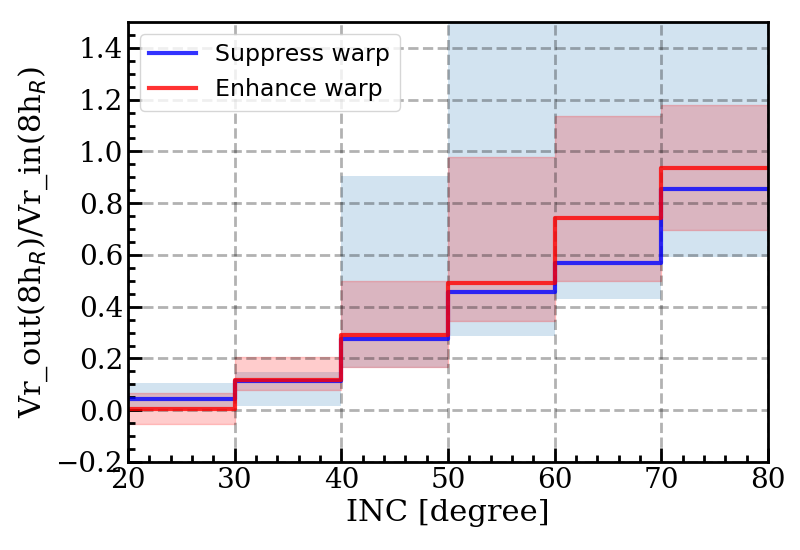,clip=true, width=0.33\textwidth}
  \end{center}
  \caption{  The same as Figure \ref{fig:10}, but for the results of using the same  weighting scheme as in \cite{DiTeodoro-21} during the fittings.  Specifically, we fit the geometry of the warps and rotation velocities by weighting the pixels with $cos^2(\theta)$ in the first steps, and then fit the radial velocity by weighting the pixels with $sin^2(\theta)$, where the $\theta$ is the angle to the kinematic major axis.  }
  \label{fig:13}
\end{figure*}

The left and middle panels of Figure \ref{fig:9} shows the MAE (defined as above) as a function of radius for all the 200 mock galaxies, using uniform weighting of the pixels.  We show the case of warp-suppression in the left panel and warp-enhancement in the middle panel. In the left and middle panels, the MAE of 200 mock galaxies are displayed separated into three subsamples, according to INC. 

As shown, the MAE of mock galaxies remain rather low for the vast majority of the population ($<$15 km s$^{-1}$), with only a few outliers.  Strikingly, the outliers with catastrophic MAE tend to have large INC.  We have examined the velocity fields of these outliers, and find that large INC more likely lead to the situations in which motions of gas at different radii are superposed at the same location in the LOS velocity fields. This increases the difficulty of the fittings, and is probably the reason for the catastrophic modeling of these mock galaxies. 

We have also examined that the MAE depends on the ${\tt vr\_scale}$ for the full mock sample as expected, and shows nearly no dependence on $\Delta$PA. As was pointed out in Section \ref{sec:4.2}, galaxies with higher INC tend to have higher LOS velocities, and this may contribute to the dependence of MAE on INC.  To eliminate this effect, we have therefore examined the normalized MAE by the maximum LOS velocity (with respect to the systematic velocity), and found that the normalized MAE indeed still depends on the INC as suggested from Section \ref{sec:3}.  

To quantify the dependence on INC, we show the median MAE(8$h_{\rm R}$) as a function of input inclination in the right panel of Figure \ref{fig:9}, for the cases of both suppressing and enhancing warps. The shaded regions show the 16-84\% percentiles of the mock galaxies.   As shown, the median MAE(8$h_{\rm R}$) increases strongly with INC, from $\sim$2 km s$^{-1}$ at INC of 30 degree to $\sim$12 km s$^{-1}$ at INC of 70 degree.  

The left and middle panels of Figure \ref{fig:10} show the extracted radial velocity (normalized by the input radial velocity at $\sim8h_{\rm R}$), as a function of radius for the mock galaxies with ${\tt vr\_scale}>0.3$.  Again, Figure \ref{fig:10} is for uniform weighting of all pixels.  As can be seen, overall the recovered radial velocities ${\tt Vr\_out(r)}$ are much smaller than the real (input) ones for the vast majority of mock galaxies. The ${\tt Vr\_out(r)}$/${\tt Vr\_in(8h_{\rm R})}$ show a strong dependence on the INC of the galaxies in that highly inclined galaxies have a higher  ${\tt Vr\_out(r)}$/${\tt Vr\_in(8h_{\rm R})}$.  

In a similar way as in Figure \ref{fig:9}, we show the median ${\tt Vr\_out(8h_{\rm R})}$/${\tt Vr\_in(8h_{\rm R})}$ as a function of INC in the right panel of Figure \ref{fig:10}, with the shaded regions showing the 16-84\% percentiles of the mock galaxies.  The median ${\tt Vr\_out(8h_{\rm R})}$/${\tt Vr\_in(8h_{\rm R})}$ also increases strongly with INC, from 0.04 at INC of 30 degree to above 0.6 at INC$>$70 degree. Based on Figure \ref{fig:9} and Figure \ref{fig:10}, it appears that the scatters of MAE(8$h_{\rm R}$) and ${\tt Vr\_out(8h_{\rm R})}$/${\tt Vr\_in(8h_{\rm R})}$ are very large for galaxies with INC$>$50 degree for the case of suppressing warps. This is because there are more bad fits for highly inclined galaxies, with catastrophic residuals, for the cases of suppressing warps than enhancing warps. 
Based on Figure \ref{fig:10}, we conclude that, uniform weighting of pixels, the radial motion is hidden by more than 55\% when fitting composite galaxies with pure warp models for low-to-intermediate inclination galaxies (INC$<$70 degree). 

For comparison, we show the recovered radial velocities adopting the same weighting scheme as in \cite{DiTeodoro-21}, in Figure \ref{fig:13}.  Comparing Figure \ref{fig:13} with Figure \ref{fig:10}, we do find that the radail velocities are better recovered with the weighting scheme of \cite{DiTeodoro-21}, especially for highly-included galaxies.  However,  even with the weighting scheme, the radial motion is significantly hidden by $\sim$50\% for galaxies with INC of 50-60 degree. 

We note that we do not recover the full input radial velocities, as in the tests of \cite{DiTeodoro-21} despite following a similar fitting procedure, indicating an apparent inconsistency.  This may be due to the nature of of input radial motions.   We input a radially increasing inflow velocity, which therefore systematically modifies the kinematic major and minor axes (see Section \ref{sec:3}). We believe that \cite{DiTeodoro-21} used more complex patterns of radial velocities, with both inflows and outflows at different radii.  These could also be expected to produce a smaller systematic effect on the distortions of the kinematic major and minor axes.

A possible approach to decompose radial motions and warped disk geometries is to use the column density map of gas, which independently gives the geometric parameters of the disk.  However, as discussed above, the gas distribution is usually clumpy, along with the presence of spiral arms, which makes this method not widely applicable.  In addition, in the present work we ignore the possible cases that the gas at different radius orbits to slightly different centers, the possible elliptical orbits of gas motion, and any variations with azimuth angles.  Including these would make the extraction of radial motion of gas more difficult.  

In Section \ref{sec:4.1}, we have shown in Figure \ref{fig:12.1} the distribution of $\Delta$PA and INC for the galaxy sample taken from \cite{DiTeodoro-21}.  In the observations, the radial changes of position angle are mostly in the range from a few up to about 30 degrees, with a mean value of 12 degrees (we exclude the two galaxies with extremely high $\Delta$PA when calculating the mean value).  

In Section \ref{sec:3.2}, we find that the radial change of kinematic major axis is generally dominated by the effect of the warps rather than the inflows, and the $\Delta$PA for the sample galaxies in \cite{DiTeodoro-21} is 12 degree. Therefore, this can provide a rough upper limit on the inflow velocity using Equation \ref{eq:9} and \ref{eq:12}: 
\begin{equation} \label{eq:13}
    v_r / v_{\phi} < {\tt tan}(12^{\circ}) \simeq 0.21. 
\end{equation}
In other words, Equation \ref{eq:13} requires $v_r<$45 km s$^{-1}$ if the radial motion is not to dominate the casue of the radial change of position angle. 

Although the upper limits to the radial velocity appear to be lower than the predictions, it is important to appreciate that this does not invalidate our modified accretion disk model. The modified accretion disk is only expected or required to work at those radii within which most star formation happens. This is because the actual star-formation profile is the key input to the model and at radii where this is not constrained, the required inflow velocities are correspondingly uncertain.  It should be noted that more than 90\% of star formation occurs within 4 disk scale-lengths in an exponential SF disk. As shown in Figure \ref{fig:1}, the maximum radial inflow velocity at this radius is only $\sim$ 20 km s$^{-1}$, which can be well hidden within the warps.

\section{Summary and Conclusion} \label{sec:5}

% --- results ---- 
%1. the inflow velocity predicted by modified accretion disk 
Hydrodynamical simulations suggest that the inflow of gas onto galaxies is almost co-planar and broadly co-rotating with the gas disk, regardless of its thermal history, and the outflowing gas potentially leaves the disk from two sides of the gas disk \citep{Keres-05, Stewart-11, Danovich-15, Stewart-17, Stern-20, Peroux-20, Trapp-21, Hafen-22, Gurvich-22}. Motivated by this, we constructed a modified accretion disk (MAD) model to describe the formation of gas disks for SF galaxies \citep{Wang-22, Wang-22b}.  Similar to but different from the accretion disk of compact objects, the accretion disk of galactic gas disk is ``leaky" because of the removal of mass from the inflow due to star formation on the disk and the associated outflows.  In earlier papers of this series, we found that the MAD model can very well recover the observed profiles of gas-phase metallicity in the disk, and also proposed that MRI is the most plausible source of viscosity for driving the co-planar inflow and, under certain plausible circumstances, will produce an exponential profile of the SF disk. 

Observationally, however, there is a noticeable lack of evidence for significant co-planar inflows in disks.  In the present work, we therefore firm up the prediction of the inflow velocity in the MAD model, investigate the effects of the required radial motion on the observed 2-d LOS velocity field, and explore the possible origin for this inconsistency, and in particular address the question whether the required flows could be hidden within the kinematic signatures of warped disks.

The main results of this analysis are as follows. 
\begin{itemize}

\item After adopting a realistic gas surface density from observations, we find that the velocity of the co-planar inflow is likely to be only a few km s$^{-1}$ within two disk scalelengths. It gradually increases with radius and can reach to around 50-100 km s$^{-1}$ at the very outer disk.  This is broadly consistent with the results from the hydrodynamical simulations \citep{Trapp-21, SWang-22, Hafen-22}.  

\item  The radial inflow will distort the observed 2-d LOS velocity field from that expected for pure rotation.  Both the kinematic major and minor axes are distorted from the geometric ones, even though the effect of the inflow is by definition zero along the major geometric axis. This signature of inflow resembles the effect of a spatial warp of the disk (see Figure \ref{fig:3}).  %indicating a loose degeneracy between these two effects.

\item  In detail, it is found that the distortion of the velocity field by the radial motion in the region of the minor axis is larger than that in the region of the major axis, whereas for a warped disk these are in principle the same. But this difference depends quite strongly on the inclination of the disk in the sense that the differences between major and minor axis become more significant with increasing INC and vanishes as the inclination approaches zero (see Equation \ref{eq:9} and \ref{eq:12}).  

\item The twisting of the iso-velocity contours that characterises the distortion of the 2-d velocity field by gas inflow depends on the rotation sense of the galaxy but the similar pattern that arises due to warps should not (at least in a symmetric universe).   Assuming that the spiral arms in galaxies are ``trailing", the twists of the kinematic distortion due to inflow should therefore be in the same sense (``S" or ``Z") as that of the spiral arms.  In contrast, the broadly similar kinematic distortions arising from spatial warps of the disks should be uncorrelated with the sense of the spiral arms.    By visually inspecting the HI velocity fields and visible-light images for a sample of nearby galaxies, no correlation is found. This suggests that the distortions of the kinematic major axis and the iso-velocity contours is {\it mainly} due to a warped geometry of the HI disk, rather than the radial motion.

\item  However, when we construct the mock HI datacubes for galaxies with both inflows and warped disks, and then force fit these 3d datacubes with pure-warped disk models, we find that the latter can very well match the composite velocity fields for most mock galaxies with INC$<$70 degree. The goodness of fits mainly depends on the INC.  The radial motions that are present in these mock galaxies can be hidden by 55\% or more within the warped disk for low-to-intermediate inclined galaxies (INC$<$70 degree), when we use a uniform weighting of pixels in the fits.  Adopting the azimuthal weighting schemes of \cite{DiTeodoro-21} results in an improvement in the recovered radial motions, while they are still far from the true radial motion for low-to-intermediate inclined disks. 
%By examining the map of velocity residuals, we find that $\sim$85\% of mock galaxies have MAE in the outer parts (at 8$h_{\rm R}$) less than 10 km s$^{-1}$ for radial velocities of order 90 km s$^{-1}$. 

This indicates that the substantial radial inflow can hidden when modelling the 3d datacubes with pure warp model, at least for low-to-intermediate inclined disks. % We also discuss the practical difficulties of extracting radial motion for highly inclined disks. 

\end{itemize}

% --- discussion ---- 
% 1. the  of warped disk with radial motion 

We stress that radial motions can distort both the kinematic major and minor axes in 2-d projected velocity fields, especially for weakly inclined disks. 
This makes the extraction of inflow velocity to be extremely difficult based on the LOS velocity fields.  
Previous studies have usually attributed the radial change of the kinematic major axis only to the warped disk \citep[e.g.][]{Jozsa-07, de-Blok-08, Kamphuis-15, Oh-18, Schmidt-16, DiTeodoro-21}. We have shown that this will significantly underestimate the amplitude of the inflow velocity.  We also stress that due to the similarity of the above two effects for weakly inclined disks, the signatures of radial motion are potentially to be found in the more highly inclined disks. 

Combining the result from Section \ref{sec:3.2} that warps, rather than inflows, appear to be the dominant effect distorting the kinematic position angles, we provide a rough upper limit of the inflow velocities of $v_r/v_{\phi}<0.21$.  This upper limit appears to be lower than the one predicted in the hydrodynamical simulations and by our own MAD model at large radii. However, as pointed out in the introduction, this inconsistency can be due to several things. First, the gas with significant radial motion at the very outskirt of gas disk may not be neutral.   Second, the spatial coverage of the current HI surveys may still not be large enough to see significant radial motions.   
%However, we note that the predicted radial velocities increase strongly with radius, and would only be expected to dominate at large radii ($>6h_{\rm R}$).  
Mapping the HI gas to larger radii with better sensitivity is needed before drawing a firm conclusion.  The HI survey by the Square Kilometre Array will likely provide the data to do this.   Finally it should be appreciated that our own viscous MAD disk model is much better defined in the central parts of the disk, where most star-formation occurs, than in the outer regions, and indeed the outer boundary of the accretion disk in this model is not, and does not need to be, well-defined.  

\acknowledgments

We thank the anonymous referee for their constructive comments and suggestions, which greatly improved the paper. We also thank Gabriele Pezzulli, Bärbel S. Koribalski, Lin Lin and Enrico M. Di Teodoro for useful discussions during the course of this work. 

\bibliography{rewritebib.bib}
\end{document}